\documentclass[pra,preprintnumbers,nobalancelastpage,twocolumn,superscriptaddress,showpacs,nofootinbib]{revtex4-1}

\usepackage{amsmath, amssymb}
\usepackage{color}
\usepackage{graphicx}
\usepackage{subfigure}
\usepackage{xifthen}
\usepackage{booktabs}
\usepackage{ulem}
\usepackage[colorlinks=true, citecolor=blue, linkcolor=blue]{hyperref}
\usepackage{braket}

\graphicspath{{figures/}}

\def\vrhot{\vv{\rho}\,^{(t)}}

\def\spe{\mathcal{S}}
\def\sites{L}
\def\pnum{M}

\newcommand{\be}{\begin{equation}}
\newcommand{\ee}{\end{equation}}
\newcommand{\ba}{\begin{aligned}}
\newcommand{\ea}{\end{aligned}}

\newcommand{\Tr}{\operatorname{Tr}}

\newcommand{\rhohbase}{\rho^{(h)}}
\newcommand{\rhotbase}{\rho^{(t)}}
\newcommand{\rhoh}[1]{\rhohbase_{#1}}
\newcommand{\rhot}[1]{\rhotbase_{#1}}
\newcommand{\mm}[1]{{\boldsymbol #1}}
\newcommand{\vv}[1]{{\vec #1}}

\newcommand{\titleinfo}{Energy transport in an integrable parafermionic chain via generalized hydrodynamics}

\begin{document}

\title{\titleinfo}

\author{Leonardo Mazza}
\affiliation{D\'epartement de Physique, \'Ecole Normale Sup\'erieure / PSL Research University, CNRS, 24 rue Lhomond, F-75005 Paris, France}

\author{Jacopo Viti}
\affiliation{ECT \& Instituto Internacional de F\'isica, UFRN, Campos  Universit\'ario,  Lagoa  Nova  59078-970  Natal,  Brazil}

\author{Matteo Carrega}
\affiliation{NEST, Istituto Nanoscienze-CNR and Scuola Normale Superiore, Piazza San Silvestro 12, 56127 Pisa, Italy}

\author{Davide Rossini}
\affiliation{Dipartimento di Fisica, Universit\`a di Pisa and INFN, Largo Pontecorvo 3, I-56127 Pisa, Italy}

\author{Andrea De Luca}
\affiliation{The Rudolf Peierls Centre for Theoretical Physics, Oxford University, Oxford, OX1 3NP, United Kingdom}

\begin{abstract}
  We study energy transport in the integrable $\mathbb Z_3$ parafermionic chain using the partitioning protocol.
  By exploiting the Bethe-ansatz solution for the thermodynamics of the system, we develop a generalized hydrodynamic description
  of the non-equilibrium steady states, which we benchmark using numerical simulations based on matrix product states.
  The model features a low-energy conformal limit with central charge $c=4/5$, which affects the low-temperature energy current,
  as we explicitly show.
  Moreover, we exploit that, for energies close to the maximally excited state, the system is also critical and described by a conformal field theory with $c=1$. By considering the two halves prepared at two temperatures both low in value but opposite in sign, we are able to investigate in an exact and controlled way the junction between two conformal field theories with different central charges.
  Notwithstanding the absence of global conformal invariance, we find results that approximate to a high degree
  those of out-of-equilibrium conformal field theories.
  Our study extends the generalized hydrodynamics to a novel framework, where it can be profitably used
  for exploring new physical phenomena.
\end{abstract}

\maketitle

\section{Introduction}

Understanding and controlling energy transport is a theme of fundamental importance, and especially in one-dimensional quantum physics.
The recent groundbreaking experimental progress with cold atomic gases has spurred interest in the study
of coherent quantum evolution~\cite{kww-06, HLFS07, bdz-08, gklk-12, fse-13, lgkr-13, glms-14, langen-15},
with a particular emphasis to the transport dynamics~\cite{brantut-12, brantut-13}.
From a theoretical point of view, the emergence of an anomalous ballistic behavior that defies the diffusive one expected from Fourier's law,
and its interplay with integrability, have been recently widely
inspected~\cite{HHCB03, nonequilibrium2013Karrasch, LHMH11, KaMH14, BDLSNature, nonequilibrium2016Vasseur, BDVR16, SHZB16}.
For integrable models~\cite{takahashi, booksIntegrability, gaudin}, a key result of this research endeavour has been the development of a
\textit{generalized hydrodynamics} (GHD) theory, describing the spatial arrangement, in the long-time limit, of the conserved
charges preserved by the dynamics~\cite{CaDY16, BCDF16}.
Such approach is based on a compact continuity equation, which accounts for the flow of all conserved quantities between macroscopic subparts
of the sample, which are supposed to be locally equilibrated to a generalized Gibbs
ensemble~\cite{rdyo-07, fagottiIsingPRL, fagottiIsingJSTAT, Pozsgay14, EsFa16, IlievskiPRL, IlievskiJSTAT}.

The current theoretical paradigm for the study of transport in isolated quantum evolution is represented
by the so-called \textit{partitioning protocol} (PP)~\cite{partitioning}. 
Within this description, transport can be studied as a
{\it local} quantum quench~\cite{CaCa07, pssv-11, delucaFlux, JMSJD11, BeFa16, DeLucaAlvise, PBPD17, calzona1, calzona2}, where the post-quench Hamiltonian differs
from the initial one only in a finite region of space. 
Specifically, at the beginning, two decoupled semi-infinite chains are initialized in two different conditions, characterized, e.g.,
by different temperatures or chemical potentials. Subsequently, they are joined together and let evolve in time.
Depending on the initial condition, different forms of transport can be inspected, e.g., of energy or of particles. 
Rigorous results have been derived in this setting~\cite{AsBa06, Mint11,MiSo13, Doyo15},
but a further step was represented by exact calculations in the framework of conformal field theories (CFT)~\cite{BeDo12,BeDo15, BeDo16, BeDo16Review}.
The latter were sustained by calculations in free theories~\cite{PlKa07, DVBD13, CoKa14, CoMa14, DeMV15, VSDH16, ADSV16, F16, Korm17, PeGa17},
numerical simulations~\cite{LaMi10, SaMi13, KaMH14, BDVR16}, and approximate approaches~\cite{DVMR14, CCDH14, Zoto16}.
More recently, by applying the PP to integrable models, the GHD has been shown to exactly reproduce the long-time dynamics~\cite{CaDY16, BCDF16},
leading to the discovery of a remarkable number of results, both in the
quantum~\cite{DeLuca16, DoYo16, IlDe17, BVKM17, DDKY17, DoYC17, DoSp17-2,PDCBF17, DoSY17, nullfagotti, IlDeHubbard, CDV18}
and in the classical limit~\cite{delu2016, bastianellodoyon2017, DoSp17}.
However, only two paradigmatic models have been considered so far: the XXZ spin-1/2 chain, and the Lieb-Liniger bosonic model.
In this context, hydrodynamic results have been always validated by numerical simulations obtained with matrix product states (MPS)~\cite{Schollwock-MPS}.

Here we investigate, for the first time, energy transport in the $\mathbb Z_3$-integrable parafermionic
chain~\cite{Albertini1993, AlbertiniTherm, Kedem, KedemMcCoy}, through a hydrodynamic approach.
This model has recently resurged to a widespread attention because of several realistic proposals for an experimental implementation
in hybrid superconductor-semiconductor devices~\cite{lindner2012, clarke2013, cheng2012, vaezi2013, alicea2016}. 
Adding up to previous works that discussed the richness of its thermodynamics, when compared to analogous fermionic chains~\cite{jermyn2014, zhuang2015},
we show that remarkable novel phenomena emerge also in the out-of-equilibrium framework.
An experimental verification of our findings stands as an intriguing and challenging perspective.

The development of a GHD for this model is based on its Bethe-ansatz (BA) solution, which differs from that of the XXZ chain
in the following aspects. First, this parafermionic chain admits composed excitations which are not of string form~\cite{takahashi, Albertini1993}.
Second, the BA equations
impose microscopic constraints on the thermodynamic description, for which not all the excitations are actually independent.
The hydrodynamic solution presented here, together with its careful numerical validation by means of MPS, stands as a first non-trivial extension
of such general phenomenological concepts to a qualitatively different situation.
We also show that this model allows for a further non-trivial verification of the universal law for low-temperature energy transport
in a CFT, derived by Bernard and Doyon (BD)~\cite{BeDo12, BeDo15}. According to such result, the steady-state energy current should only depend
on the central charge $c$. So far, all verifications dealt with theories with $c=1$ or $c = 1/2$, corresponding to a free bosonic and fermionic theory respectively; here, the low-temperature scaling limit is a CFT with $c=4/5$ and as such our is the first check of BD law in a truly interacting theory.

Additionally, even if the low-energy limit of the Hamiltonian $\hat H$ of our model is described by a CFT with $c=4/5$,
the Hamiltonian $-\hat H$ has a low-energy conformal description with $c=1$.
The study of energy transport between models with different properties and scaling limits is an exciting possibility
for which only limited information has been discovered so far~\cite{SoCa08, vitiimpurity}. By joining together two chains prepared at opposite temperatures,
i.e.~positive on one side and negative on the other one, we develop a GHD description of a transport protocol where, effectively,
a CFT with $c=4/5$ and a CFT with $c=1$ have been joined together. Although quasi-particles are not described by a CFT,
the model features properties that approximate it to a large degree, and yield interesting physical properties.

This article is organized as follows.
In Sec~\ref{Sec:Model} we present the model and review its equilibrium BA solution. 
In Sec.~\ref{Sec:Transport} we introduce the PP and develop the hydrodynamic description, that we benchmark with numerical MPS simulations.
In Sec.~\ref{Sec:LowT} we study the low-temperature energy transport, and verify the BD law.
In Sec.~\ref{Sec:OppT} we study transport in the opposite temperature regime.
Our conclusions are drawn in Sec.~\ref{Sec:Conc}.

\section{Model}\label{Sec:Model}

\subsection{Hamiltonian}

Let us consider a one-dimensional chain of $\mathbb Z_3$-parafermions, of length $L$.
Each lattice site $\ell = 1, \ldots, L$ is associated to two parafermionic operators,
$\hat \gamma_{2\ell-1}$ and $\hat \gamma_{2 \ell}$, satisfying 
$\hat \gamma_k^3 = 1$, $\hat \gamma_{k}^\dagger = \hat \gamma_k^2$, and 
\begin{equation}
\hat \gamma_k \hat \gamma_m = \omega \, \hat \gamma_m \hat \gamma_k \;\; (k<m), \quad \mbox{with } \; \omega = e^{2 \pi \imath /3}.
\end{equation}
The system Hamiltonian is given by: 
\begin{equation}
  \label{Eq:Ham:Pf}
  \hat H_0 = -J \sum_\ell \left[ \hat \gamma_{2\ell-1} \hat \gamma_{2\ell} + \hat \gamma_{2\ell} \hat \gamma_{2\ell+1} + {\rm H.c.} \right],
\end{equation}
where $J>0$ fixes the energy scale, and we have adopted units of $\hbar = k_B = 1$~\cite{alicea2016}.

It is now convenient to introduce the Fradkin-Kadanoff transformation, which unitarily maps the parafermionic operators
into commuting $\mathbb Z_3$ variables~\cite{fradkin1980}.
To this purpose, we define the operators $\hat \tau_\ell$ and $\hat \sigma_\ell$, such that they satisfy the following algebra:
$\hat \sigma_\ell \hat \tau_\ell = \omega \, \hat \tau_\ell \hat \sigma_\ell$ and are otherwise commuting.
These operators are represented in the single-site Hilbert space by the following matrices:
\begin{equation}
  \tau = \begin{pmatrix}
    0 & 0 & 1\\
    1 & 0 & 0\\
    0 & 1 & 0
  \end{pmatrix} \;, \qquad
  \sigma = \begin{pmatrix}
    1 & 0 & 0\\
    0 & \omega & 0\\
    0 & 0 & \omega^2
  \end{pmatrix} .
\end{equation} 
The mapping reads:
\begin{equation}
  \label{eq:map_FK}
 \hat \gamma_{2\ell-1}  = \omega \hat \sigma_\ell^\dagger \hat \tau_\ell^\dagger \prod_{k > \ell} \hat \tau_k^\dagger;
 \quad  \quad
 \hat \gamma_{2 \ell-1}  = \hat \sigma_\ell^\dagger \prod_{k > \ell} \hat \tau_k^\dagger.
\end{equation}
The Hamiltonian $\hat H_0$ in Eq.~\eqref{Eq:Ham:Pf} is thus mapped into:
\begin{equation}
\label{ham3}
\hat H = \sum_{\ell} \hat h_\ell \;, \qquad
\hat h_\ell = -J \left[ \hat \tau_\ell + \hat \sigma_\ell^\dag \hat \sigma_{\ell+1} + {\rm H.c.} \right], 
\end{equation}
and in the following we will make explicit reference to it.
We stress that the transformation~\eqref{eq:map_FK} is unitary, and we will only consider operators whose locality
is preserved by the mapping; as such, our conclusions apply also to the parafermionic version of the model.

\subsection{Bethe-Ansatz formulation}

The Hamiltonian in Eq.~\eqref{ham3} is integrable via a standard BA technique, and in the following we briefly summarize its solution, originally presented in Refs.~\cite{Albertini1993, AlbertiniTherm, Kedem}.
Let us start by mentioning that the model is invariant under the $\mathbb{Z}_3$ symmetry 
\begin{equation}
  [\hat H, \hat U] = 0\;, \qquad \hat U = \prod_{j=1}^\sites \hat \tau_j \;.
\end{equation}
Since $\hat U^3 = \mathbf{1}$, the eigenvalues of $\hat U$ are $e^{2\imath \pi Q/3 }$, with $Q = \{-1, 0, 1\}$, corresponding to three symmetry sectors. 

\subsubsection{Single-particle spectrum}

We are interested in the spectrum of Eq.~\eqref{ham3} in the thermodynamic limit $L \to \infty$.
Let us first consider finite $L$ values and periodic boundary conditions,
such that exact eigenstates at finite size are associated to sets of rapidities (or roots) $\{ \lambda_1,\ldots, \lambda_\pnum\}$,
solutions of the BA equations~\cite{Albertini1993, AlbertiniTherm, Kedem}
\begin{equation}
\label{BAE}
\left[ 
\frac{\sinh ( \frac{\imath\pi}{12} - \lambda_j)}
{\sinh ( \frac{\imath\pi}{12} + \lambda_j)}
\right]^{2\sites} \! = (-1)^{\sites+1} \prod_{k=1}^{\pnum} \frac{\sinh\big[\frac{\imath\pi}{3}  - (\lambda_j - \lambda_k)\big]}{\sinh \big[\frac{\imath\pi}{3}  - (\lambda_j - \lambda_k)\big]} ,
\end{equation}
where the number $\pnum$ of rapitidies is fixed by the sector of the $\mathbb{Z}_3$ symmetry $Q$: $\pnum = 2(\sites - |Q|)$.
Finding a solution of Eqs.~\eqref{BAE} for an arbitrary $\sites$ is generally hard.
Nonetheless, the structure of its solutions considerably simplifies for $\sites \to \infty$, where the rapidities can be grouped according to the arrangement of their imaginary parts. There are five classes of rapidities labelled by $\spe = \{ \mbox{(a)}, \mbox{(b)},\mbox{(c)},\mbox{(d)},\mbox{(e)}\}$ whose properties are given in Tab.~\ref{classes} (see Ref.~\cite{Albertini1993} for further details).
\begin{table}
\begin{tabular}{ | c | l | c | c | c|}
\hline
  Label ($\mu$) & Roots & $n_\mu$ & $\upsilon_\mu$ & $\sigma_\mu$\\
  \hline
  (a) & $\lambda_{k,a}$ & 1 & 1 & 1\\
  (b) & $\lambda_{k,b} + \frac{\imath \pi}{2}$  & 1 & -1 & 1\\
  (c) & $\lambda_{k,c} \pm \frac{\imath \pi}{3}$ & 2 & 1 & -1\\
  (d) & $\lambda_{k,d} + \frac{\imath \pi}{2} \pm \frac{\imath \pi}{3}$ & 2 & -1 &-1\\
  (e) & $\lambda_{k,e}  \pm \frac{\imath \pi}{4}$ & 2 & -- & -1\\
  \hline
\end{tabular}
\caption{Spectrum of single particles in the BA solution of Eqs.~\eqref{ham3}, for $L\to \infty$. 
  In this notation $\lambda_{k,\mu}$ is a real number, with $k =1,\ldots, M_\mu$.
  Here $n_\mu$ indicates the length, $\upsilon_\mu$ the parity, and $\sigma_\mu$ the sign of each rapidity.}
\label{classes}
\end{table}
For each label $\mu \in \spe$, there are $M_\mu$ distinct real parts $\{\lambda_{k, \mu}\}_{k=1}^{M_\mu}$ and for each real part there are $n_\mu$ rapidities which differ in their imaginary part. In other words, the initial set of roots is splitted as
\begin{equation}
\label{stringsplitting}
M = \sum_{\mu \in \mathcal{S}} n_\mu M_\mu \;.
\end{equation}
The label in $\spe$ can be associated with five particle types which constitute the single-particle spectrum of the model.
Note that the classes (c), (d), (e) are composed by two complex conjugate rapidities: they can be considered as stable bound states
composed of two elementary particles. A reader familiar with the BA formalism can recognize classes (a), (b), (c), (d) as string configurations
[(b) and (d) having a negative parity]~\cite{takahashi}.
On the other hand, class (e) is a peculiarity of this model and, more in general, of $\mathbb{Z}_N$-integrable chains~\cite{AlbertiniTherm}.

\subsubsection{Thermodynamic eigenstates and conserved quantities}

In the thermodynamic limit, each eigenstate is occupied by an extensive number of particles. For each particle type $\mu \in \spe$, the $M_\mu$ real parts become dense on the real line and are described by a density of roots  $\rho_\mu(\lambda)$:
\begin{equation}
\rho_\mu(\lambda) = \left.\lim_{\sites \to \infty} \frac{1}{L (\lambda_{k+1, \mu} - \lambda_{k, \mu})} \right|_{\lambda_{k,\mu} = \lambda}
\end{equation}
where we sorted the real parts for $\mu$-type particles according to: $\lambda_{k+1,\mu}  > \lambda_{k,\mu}$.

To describe the structure of the eigenstates in the thermodynamic limit, it is useful to draw an analogy with non-interacting fermions on a periodic lattice of the same size $L$. 
In that case, different eigenstates are realized by filling some among all the available momenta $2\pi n/L$ with $n \in \mathbb{N})$.  In the BA jargon, the available momenta are dubbed \textit{vacancies}, while the occupied/unoccupied ones are referred to as \textit{roots/holes}. When $L \to\infty$, an eigenstate is characterized by a density of roots among the available vacancies.
In a similar way, for the parafermionic model we are considering here, different eigenstates are obtained by all possible ``fillings'' 
of the real parts $\lambda_{k,\mu}$ among all the possible vacancies~\cite{takahashi}.  Since each vacancy is either filled or empty,
introducing the density of holes $\rhoh{\mu}(\lambda)$ and vacancies $\rhot{\mu}(\lambda)$, we have the relation
\begin{equation}
\rhot{\mu}(\lambda) = \rho_\mu(\lambda) + \rhoh{\mu}(\lambda).
\end{equation}
The main difference with respect to non-interacting models is that the density of vacancies is not a fixed function.  Actually, the BA equations~\eqref{BAE} translates into a functional relation between the density of roots and of vacancies:
\begin{equation}
\label{fullBA}
\mm{\sigma} \vrhot + \frac{1}{2\pi} \mm{\Theta}'  \vv{\rho}= \frac{\vv{t}\,'}{2\pi} \,.
\end{equation}
Notice that here we have introduced a shorthand notation for vectors:
$[ \vv{f} \, ]_\mu (\lambda) = f_{\mu}(\lambda)$ and for matrices:
$[ \mm{M} ]_{\mu,\nu} (\lambda, \lambda') = M_{\mu, \nu}(\lambda, \lambda')$,
with the matrix-vector multiplication defined as
\begin{equation}
\big[ \mm{M} \vv{f} \, \big]_{\mu} (\lambda) = \sum_{\nu} \int_{-\infty}^\infty {\rm d}\lambda' \, M_{\mu, \nu}(\lambda, \lambda') \, f_\nu(\lambda') \;.
\end{equation}
The prime instead indicates the derivative with respect to the rapidity, e.g., $[\vv{f}\,']_\mu  (\lambda)= df_\mu(\lambda)/d\lambda$.
Explicit forms of the matrix $\Theta_{\mu,\nu}(\lambda, \mu)$ and the vector $t_{\mu}(\lambda)$ are reported in App.~\ref{BADet}.
The matrix $\mm{\sigma}$ has elements $\sigma_{\mu,\nu}(\lambda, \lambda') = \sigma_\mu \, \delta_{\mu,\nu} \, \delta(\lambda - \lambda')$,
where the signs $\sigma_\mu \in \{-1, 1\}$ are given, for each particle type, in Tab.~\ref{classes}.
In general, a set of functions $\rhot{\mu}(\lambda)>\rho_\mu(\lambda) > 0$ solutions of \eqref{fullBA} represents a thermodynamic eigenstate of $\hat H$. For this model,
one can show~\cite{Albertini1993, AlbertiniTherm, Kedem} that Eqs.~\eqref{BAE} impose the following additional constraints for eigenstates to be physical
\begin{subequations}
\label{constraints}
\begin{align}
\rho_a (\lambda) = \rhoh{c}(\lambda) \;, & \qquad \rhoh{a} (\lambda) = \rho_{c}(\lambda) \;, \\
\rho_b (\lambda) = \rhoh{d}(\lambda) \;, & \qquad \rhoh{b} (\lambda) = \rho_{d}(\lambda) \;,
\end{align}
\end{subequations}
which have to be satisfied together with \eqref{fullBA}. 

Equivalently, a thermodynamic eigenstate can be defined in terms of the filling factors
\begin{equation}
\vartheta_\mu(\lambda) \equiv \rho_\mu(\lambda)/\rhot{\mu}(\lambda) \;.
\end{equation} 
Indeed, using \eqref{fullBA}, one can relate the filling factors $\vartheta_\mu(\lambda)$ to the corresponding root densities $\rho_\mu^{[\vartheta]}(\lambda)$ via
\begin{equation}
\label{rhofromtheta}
\vv{\rho}\,^{[\vartheta]} = \left(\mm{\sigma} \mm{\vartheta}^{-1} + \frac{1}{2\pi} \mm{\Theta}'\right)^{-1} \frac{\vv{t}'}{2\pi} ,
\end{equation}
where we introduced the diagonal matrix containing the filling factors
\begin{equation}
[\mm\vartheta]_{\mu\nu}(\lambda, \lambda') \equiv \delta_{\mu,\nu} \, \delta(\lambda - \lambda') \, \vartheta_\mu(\lambda) \;.
\end{equation}
In the following, we will denote compactly as $\ket{\vartheta}$ the thermodynamic eigenstate associated to a set of filling factors $\vartheta_\mu(\lambda)$ and root densities
 $\rho_\mu^{[\vartheta]}(\lambda)$ related one another via~\eqref{rhofromtheta}.

\subsubsection{Conserved quantities and associated currents}

Because of integrability, the Hamiltonian model~\eqref{ham3} admits an infinite number of local conserved quantities, which are sums of local densities.
Each of them can be represented as
\begin{equation}
\hat Z = \sum_\ell \hat z_{\ell} \;,
\end{equation}
with  $[\hat H, \hat Z] = 0$ and where $\hat z_{\ell}$ represents the charge density, having support on a finite number of
sites~\footnote{The presence of multiple particle types suggests that this model must admit different families of quasi-local conserved quantities~\cite{PrIl13, IlievskiPRL, IlievskiJSTAT, prosenreview, PiVC16}. Here, we simply assume that a complete set of conserved quantities can be defined without specifying their explicit construction, leaving this analysis to a future study.} around $\ell$.
To each conserved density $\hat z_\ell$ is associated a corresponding current $\hat j_{\ell}^Z$, defined via the continuity equation 
\begin{equation}
\hat j_{\ell}^Z - \hat j_{\ell-1}^Z  = \imath [ \hat H, \hat z_{\ell}] \;.
\label{Eq:Continuity:General}
\end{equation}
Since the conserved quantities commute, each state $\ket{\vartheta}$ is a simultaneous eigenstate of all of them.
Being these operators local, the corresponding eigenvalue is additive on the particle content, i.e., it takes the form
\begin{equation}
\label{chargeave}
\lim_{L\to \infty} \! \frac{\bra{\vartheta} \hat Z \ket{\vartheta}}{L} =  \bra{\vartheta} \hat z_\ell \ket{\vartheta} \! = 
\! \sum_{\mu \in \spe} \int \! {\rm d}\lambda \, \rho_\mu^{[\vartheta]}(\lambda) \, \mathfrak{z}_{\mu}(\lambda).
\end{equation}
Here $\mathfrak{z}_{\mu}(\lambda)$ is the single-particle eigenvalue which quantifies the contribution of a particle of type $\mu$ and rapidity $\lambda$ to the charge $\hat Z$. For instance, for the Hamiltonian and the momentum, the single-particle eigenvalue respectively takes the form
\begin{subequations}
\begin{align}
&\mathfrak{e}_{\mu}(\lambda) = \frac{\sqrt{3}}{8} (1 + \delta_{\mu, e}) t'_\mu(\lambda) = \big[ \mm{S} \vv{t}\,' \big]_{\mu}(\lambda)\\
&\vec{\mathfrak{p}} = -\frac{1}{2}\mm{S} \vv{t} \;, 
\end{align}
\end{subequations}
where we introduced the diagonal matrix
\begin{equation}
S_{\mu,\nu}(\lambda, \mu) = \delta_{\mu,\nu} \, ( 1 + \delta_{\mu,e}) \, \delta(\lambda - \mu).
\end{equation}
Via Eq.~\eqref{chargeave}, the root densities $\rho_\mu(\lambda)$ are in one-to-one correspondence with a complete set of conserved quantities. 
Therefore, the state $\ket{\vartheta}$ can be equivalently considered as a microcanonical representative 
of the generalized Gibbs ensemble~\cite{rdyo-07, ViRi16, quenchactionEsslerCaux, quenchactionreview}.

Even though the state $\ket{\vartheta}$ is not an eigenstate of the currents, a formula similar to \eqref{chargeave}
holds for their expectation value~\cite{CaDY16, BCDF16}
\begin{equation}
  \label{currave}
  \bra{\vartheta} \hat j_\ell^Z \ket{\vartheta} =
  \sum_{\mu \in \spe} \int {\rm d}\lambda \, v_\mu^{[\vartheta]} (\lambda) \, \rho_\mu(\lambda) \, \mathfrak{z}_{\mu}(\lambda) \;.
\end{equation}
The function $v_\mu^{[\vartheta]}(\lambda)$ describes the velocity of quasiparticles at rapidity $\lambda$ in the thermodynamic eigenstate $\ket{\vartheta}$.
For a non-interacting theory, the velocity would be simply obtained from the dispersion relation differentiating the single-particle energy
with respect to the corresponding momentum ($v_\mu  = d \mathfrak{e}_\mu / d\mathfrak{p}_\mu$). In the presence of interactions,
the single-particle energy and momentum have to be modified according to the state $\ket{\vartheta}$ (\textit{dressing}). One arrives at~\cite{bonnes14}
\begin{equation}
\label{veldef}
v_\mu^{[\vartheta]}(\lambda) = \frac{\mathcal{D}^{[\vartheta]}[\mathfrak{e}']_\mu(\lambda)}{\mathcal{D}^{[\vartheta]}[\mathfrak{p}']_\mu(\lambda)} \,,
\end{equation}
where the dressing operation $\mathcal{D}^{[\vartheta]}(f_\mu)$ in the state $\ket{\vartheta}$ acts linearly on a single-particle eigenfunction $f_\mu(\lambda)$ as 
\begin{equation}
\label{dressdef}
\vec{\mathcal{D}}^{[\vartheta]} (f) \equiv \left(1 + \frac{1}{2\pi} \mm{S} \mm{\Theta}' \mm{S}^{-1} \mm{\sigma} \mm{\vartheta}\right)^{-1} \vec f \,.
\end{equation}

\subsubsection{Thermodynamics}

The thermodynamic BA allows one to associate a representative thermodynamic eigenstate $\ket{\vartheta}$ to the thermal density matrix\footnote{Note that we use the hat to avoid confusion between the density matrix $\hat \rho$ from the root density functions $\rho_\mu(\lambda)$.}
\begin{equation}
\hat \rho = \frac{e^{- \beta \hat H}}{\mathcal{Z_\beta}} \to \ket{\vartheta} \;,
\end{equation}
where $\beta$ denotes the inverse temperature of the system, and ${\mathcal{Z_\beta}}$ the partition function.
Here we will not give details of this standard construction, which is based on minimizing the free-energy functional $\mathcal{F} = \beta \bra{\vartheta} \hat H \ket{\vartheta} - \mathcal{S}_{YY}[\vartheta]$, with $\mathcal{S}_{YY}[\vartheta]$ the Yang-Yang entropy~\cite{takahashi}. Here, we simply stress that, for this model, the minimization procedure must account for the constraints in Eq.~\eqref{constraints}, for which only three root densities $\vartheta_a(\lambda), \vartheta_b(\lambda), \vartheta_e(\lambda)$ are actually independent. Setting $\eta_\mu(\lambda) = \rho_\mu(\lambda)/\rhoh{\mu}(\lambda)$, the minimization leads to~\cite{Kedem}
\begin{subequations}
\label{TBAred}
\begin{align}
&\ln \eta_b = - K_1 \ast \ln \big[ (1 + \eta^{-1}_a)(1 + \eta_b^{-1}) \big] , \\
& \ln \eta_a = \ln \eta_b  +  \frac{3 \sqrt{3}\beta}{\cosh(6 \lambda)} , \\
&\ln \eta_e = K_2 \ast \ln \big[ (1 + \eta^{-1}_a)(1 + \eta_b^{-1}) \big] ,
\end{align}
\end{subequations}
where we omit the explicit dependence on the rapidity and we indicate the convolution as
$f \ast g = \int_{-\infty}^\infty {\rm d}\lambda' \, f(\lambda - \lambda') \, g(\lambda')$. In Eqs.~\eqref{TBAred}, we introduced the functions 
\begin{equation}
K_1(\lambda) = \frac{18\lambda}{\pi^2\sinh(6\lambda)}\;, \quad K_2(\lambda) = \frac{3}{\pi \cosh(6 \lambda)}\;;
\end{equation}
the functions $\eta_a(\lambda; \beta), \eta_b(\lambda; \beta), \eta_e(\lambda; \beta)$ can be thus easily determined numerically for arbitrary $\beta$
(either positive or negative). By employing~\eqref{fullBA} and~\eqref{constraints}, one can then obtain the whole set of filling factors
\begin{equation}
\label{varthetofrometa}
\vartheta_\mu(\lambda; \beta) = \frac{\eta_\mu(\lambda; \beta)}{1 + \eta_\mu(\lambda; \beta)}
\end{equation}
describing a thermal state at inverse temperature $\beta$.

\section{The partitioning protocol}\label{Sec:Transport}

In order to study energy transport in the system, we consider a partitioned initial state.
From the Hamiltonian density $\hat h_\ell$ introduced in \eqref{ham3},
we define the Hamiltonians relative to the left/right halves of the system
\begin{equation}
\hat H_r = \sum_{\ell > 0} \hat h_\ell \;, \qquad \hat H_l = \sum_{\ell < 0} \hat h_\ell \;.
\end{equation}
We focus on partitioned initial states, in which the two halves are at thermal equilibrium but at different temperatures,
thus exhibiting a macroscopic unbalance in the energy density, i.e.
\begin{equation}
\label{inipart}
\hat \rho_0 = \frac{e^{-\beta_l \hat H_l} \otimes e^{-\beta_r \hat H_r} }{\mathcal{Z}} \;,
\end{equation}
which is then evolved with the full Hamiltonian $\hat H$ in \eqref{ham3}.

\subsection{Generalized hydrodynamic formulation}

Despite the integrability of the model, computing the exact time evolution of $\hat \rho_0$ remains an extremely hard task.
An alternative approach is based on assuming local equilibration to a generalized microcanonical ensemble. In practice, the filling factors are promoted to space-time dependent functions $\vartheta_\mu^{(x,t)}(\lambda)$, which describe local observables around a coarse-grained space-time point $(x,t)$. Imposing the continuity equation of all conserved quantities, one can derive the GHD equation~\cite{CaDY16, BCDF16} in the form 
\begin{equation}
\label{ghdrho}
\partial_t \vartheta_\mu^{(x,t)} + v_\mu^{[\vartheta^{(x,t)}]}  \partial_x \vartheta_\mu^{(x,t)} = 0 \,,
\end{equation}
where we omitted the explicit dependence on rapidity $\lambda$ of all quantities. 
Equation~\eqref{ghdrho} has to be solved together with~\eqref{veldef}; we refer to~\cite{BVKM17}
for an analysis of efficient numerical methods to evaluate its solutions from generic initial conditions.
Once the solution $\vartheta_\mu^{(x,t)}(\lambda)$ is found,
the space-time profile of a conserved density $\hat z_\ell$ and the corresponding current $\hat j_\ell^Z$
can be obtained using~\eqref{chargeave} and~\eqref{currave}. 

For a partitioned initial state, Eq.~\eqref{ghdrho}
leads to a self-similar dynamics, where all local expectation values have a space-time profile which only depends on the ratio $x/t$. The solution can be written explicitly as 
\begin{equation}
\label{selfpart}
\vartheta_\mu^{(x, t)} (\lambda) = \begin{cases}
\vartheta_\mu(\lambda; \beta_l), & v^{[\vartheta(x,t)]}(\lambda) > x/t \\
\vartheta_\mu(\lambda; \beta_r), & v^{[\vartheta(x,t)]}(\lambda) < x/t 
\end{cases} \;, 
\end{equation}
where the thermal filling factors $\vartheta_\mu(\lambda; \beta_{l,r})$, obtained from the solutions of \eqref{TBAred} at $\beta = \beta_{l,r}$, describe the left and right initial states, in agreement with~\eqref{inipart}. 

We will consider the energy current flowing at site $\ell$, $\hat  j_\ell^{H}$, which is defined in Eq.~\eqref{Eq:Continuity:General} for $\hat z_\ell$ equal to the local Hamiltonian density $\hat  h_\ell$ in Eq.~\eqref{ham3}.
The space-time profile of the energy current is defined as
\begin{equation}
\mathcal{J}_E (x,t) = \Tr \left[ e^{-\imath \hat H t} \, \hat  j_\ell^{H} \, e^{\imath \hat H t} \hat \rho_0 \right] ,
\end{equation}
where $x \equiv \ell$.
From the solution in~\eqref{selfpart} and~\eqref{rhofromtheta}, one obtains the GHD approximation
for $\mathcal{J}_E(x,t)$ via~\eqref{currave}. 
In this approximation, the space-time profile only depends on the ray $x/t = \mbox{const.}$, i.e. 
$\mathcal{J}_{E}^{\mbox{\tiny (GHD)}}(x/t)$.
We expect the GHD approach to become more and more accurate at large times, as will become apparent
in the next section. 

\subsection{Comparison with numerical simulations}
\label{Sec:DMRG:verification}

In order to test the validity of the hydrodynamic approach developed above, we compare it to numerical simulations performed
with time-dependent MPS~\cite{Schollwock-MPS}.
The initial thermal state is obtained by purifying the system through the ancilla method; this procedure squares the dimension of the local Hilbert space.
The subsequent time-evolution is performed using a time-evolving block-decimation (TEBD) algorithm with a fourth-order Trotter expansion
of the unitary evolution operator (we fixed a time step $dt = 10^{-2}/J$), and exploiting a backward time-evolution
of the auxiliary system to optimize the growth of entanglement~\cite{KaBM12, KeKa_16}.
We consider chains up to $L=100$ with open boundary conditions, ensuring that finite-size effects are under control.
The maximum allowed bond link $m$ is specified for each simulation, and the truncation error per step is set to $10^{-10}$.

\begin{figure}[!t]
\begin{center}
 \includegraphics[width=0.95\columnwidth]{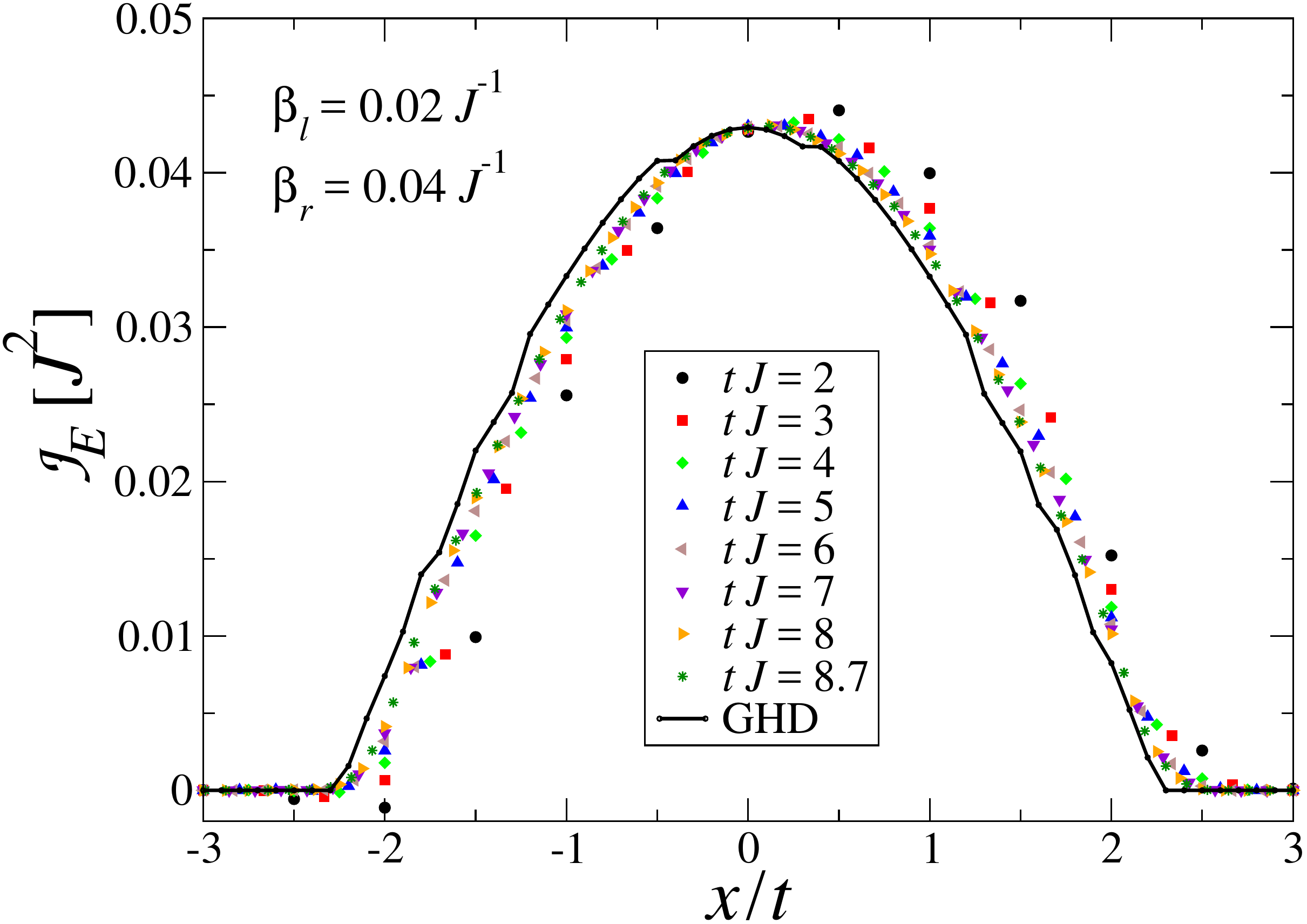}
\end{center}
\caption{Rescaled energy-current profile for $\beta_l = 0.02 J^{-1}$ and $\beta_r = 2 \beta_l$, according to the hydrodynamic prediction
  for the steady state (black solid line) and to numerical data for several finite times (colored symbols). 
  Simulations employ a maximal bond link $m = 300$ (the accuracy of the simulations is discussed in App.~\ref{App:Accuracy}).} 
 \label{Fig:Hydro:Test:Thermal:1}
\end{figure}

\begin{figure}[t]
 \begin{center}
 \includegraphics[width=\columnwidth]{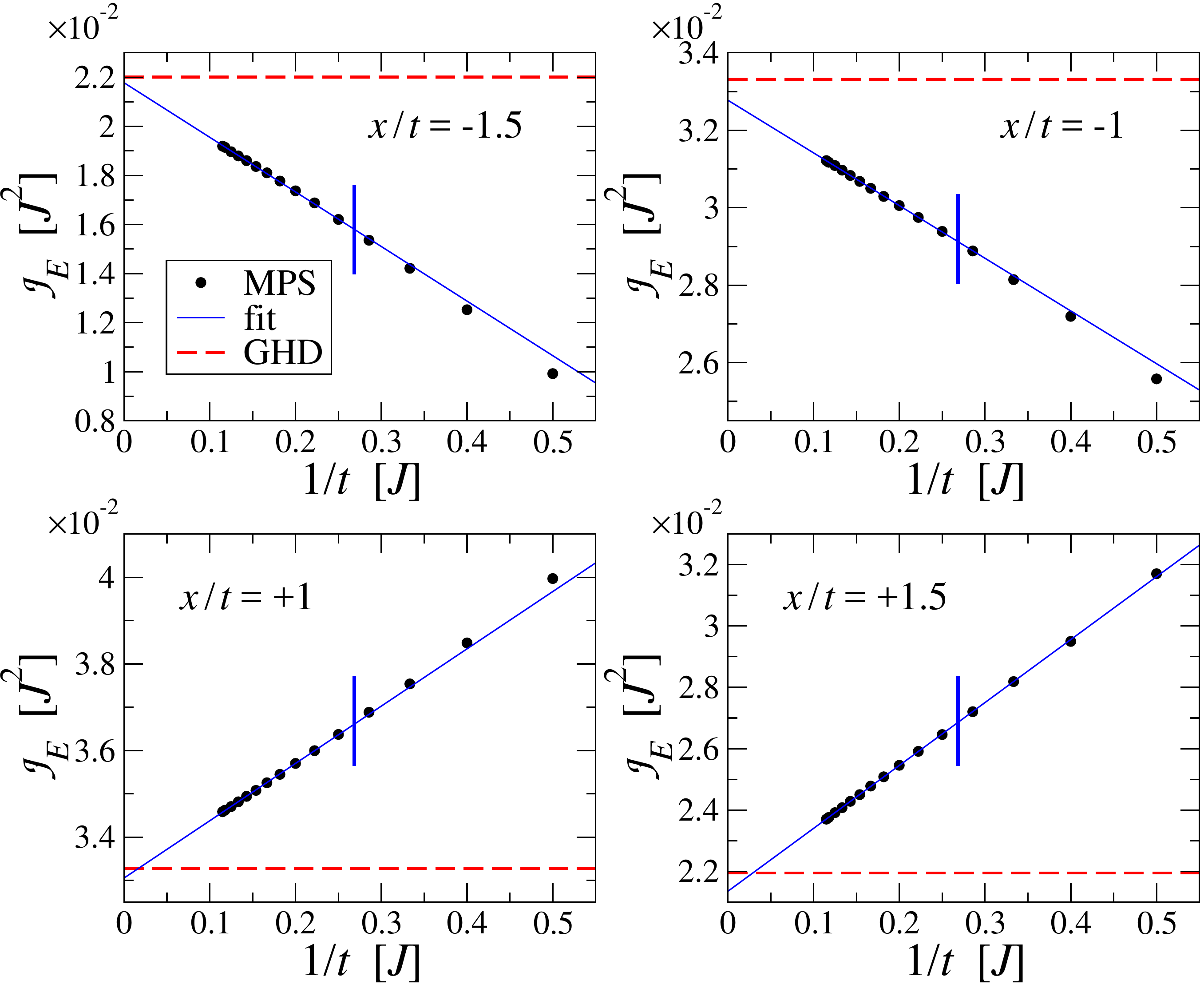}
 \end{center}
 \caption{Time-dependence of the data presented in Fig.~\ref{Fig:Hydro:Test:Thermal:1} for fixed values of $x/t = -1.5$, $-1.0$, $1.0$ and $1.5$. The blue line represents a fit to the data with a function $a_0 + a_1/t$; the fit is performed considering only data to the left of the vertical bar. The horizontal dashed red line represents the GHD prediction for the steady-state value.}
 \label{Fig:App:Hydro:Test:Thermal:2}
\end{figure}

We begin by considering the case $\beta_l = 0.02 J^{-1}$ and $\beta_r = 0.04 J^{-1}$.
In Fig.~\ref{Fig:Hydro:Test:Thermal:1} we show the rescaled profile of the energy current ${\cal J}_E$ for several times;
the hydrodynamic prediction is superimposed as a continuous black curve.
Notice that the agreement between the numerical data and the analytics is only approximate.
With respect to previous studies on the XXZ spin-1/2 chain, the problem features a greater numerical complexity, which is imputable to a larger Hilbert space
and to the absence of a U(1) symmetry; this prevents the simulations from reaching long-enough times for a direct confirmation of the theory.
To overcome this issue, we have carefully studied the time-dependence of our data for fixed values of $x/t$, as is visible
in Fig.~\ref{Fig:App:Hydro:Test:Thermal:2}.
By fitting the long-time behavior with the functional form $a_0 + a_1 / t $, we obtain asymptotic values that agree
with the hydrodynamic predictions within few percents, thus validating the GHD approach in this regime. 

\begin{figure}[t]
 \begin{center}
  \includegraphics[width=0.95\columnwidth]{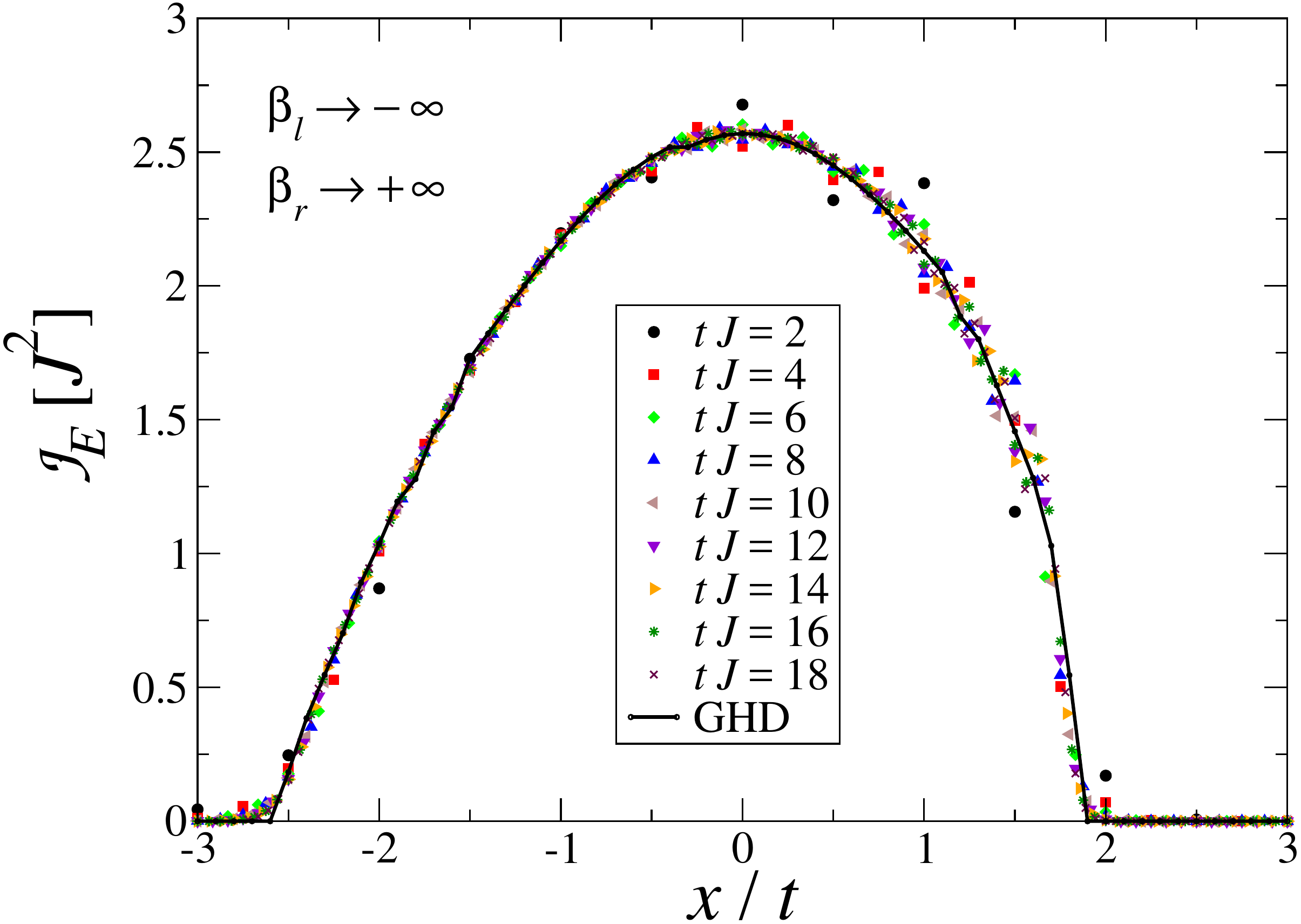}
 \end{center}
 \caption{Rescaled energy-current profile for $\beta_l \to - \infty $ and $\beta_r = \to+ \infty $:
   GHD prediction for the steady state (black solid line) and numerical data for several finite times (colored symbols). 
   Simulations employ a maximal bond link $m = 300$.}
 \label{Fig:Hydro:Test:GS}
\end{figure}

In order to circumvent the limitations due to the finite times accessible with our numerical tools,
we also consider the limiting case $\beta_l \to - \infty$ and $\beta_r \to + \infty$.
In this case, the initial state~\eqref{inipart} is a pure state and the required computational resources are significantly reduced,
enabling us to reach considerably larger times, up to $t \approx 18 J^{-1}$. 
Numerical and analytical results are shown in Fig.~\ref{Fig:Hydro:Test:GS}.
The agreement with the hydrodynamic prediction is significantly improved, as is evident from a visual comparison with Fig.~\ref{Fig:Hydro:Test:Thermal:1}.
In App.~\ref{App:Accuracy} we present an additional, more quantitative, analysis of the time-dependence of the numerical data for fixed $x/t$.

\begin{figure}
 \begin{center}
 \includegraphics[width=0.95\columnwidth]{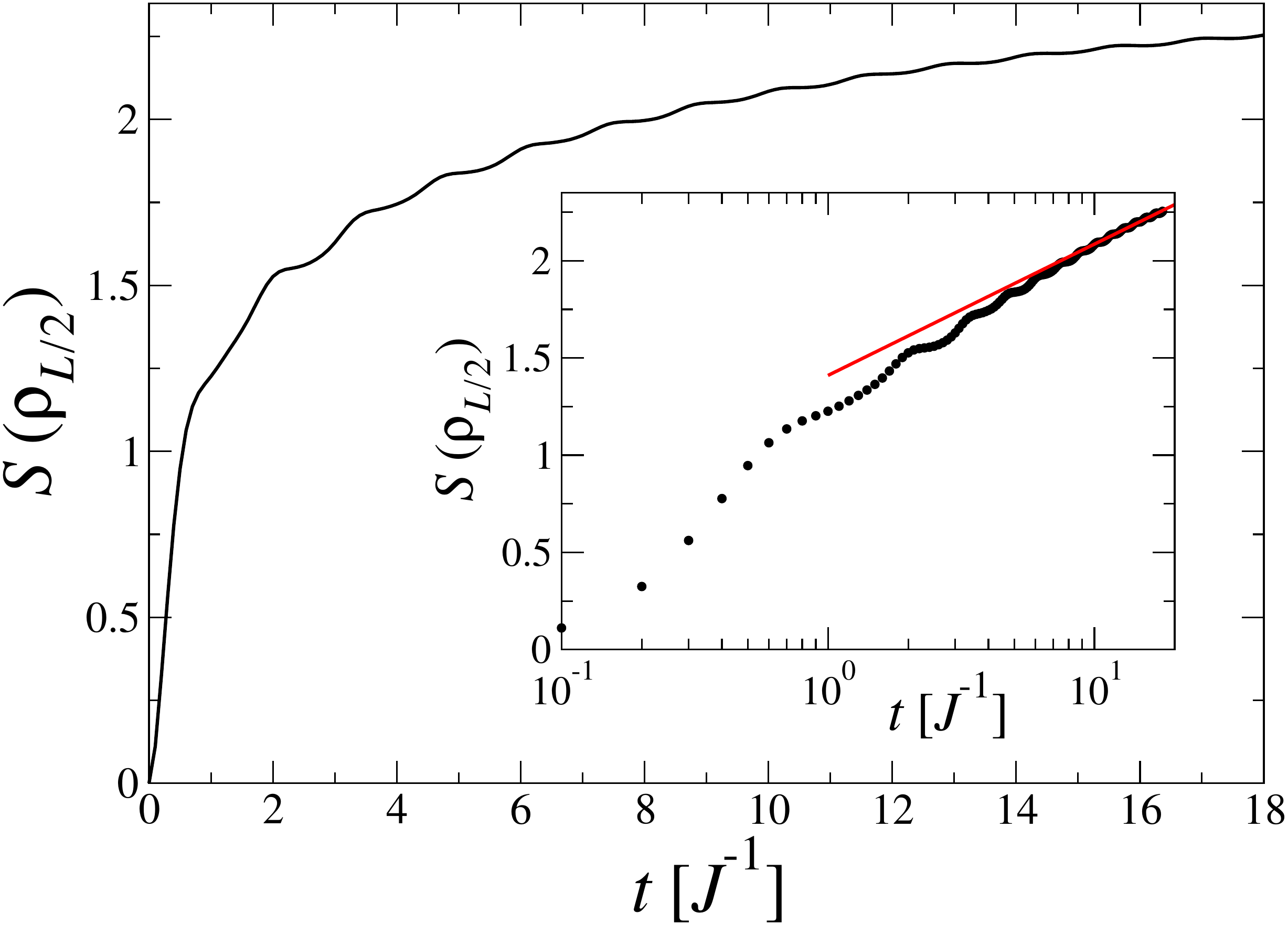}
 \end{center}
 \caption{Entanglement entropy of a bipartition of the system as a function of time, for $\beta_l \to - \infty $ and $\beta_r = \to+ \infty $.
   The inset highlights a scaling as $\sim \log (t J)$: the red curve is a fit of numerical data (black circles) for $t>10 J^{-1}$,
   which yields: $S(\rho_{L/2}) \approx 1.41 + 0.293 \times \ln(t J)$. Data correspond to the same simulations as in Fig.~\ref{Fig:Hydro:Test:GS}.}
 \label{Fig:VNE_GS}
\end{figure}

We conclude by mentioning that in the latter protocol, although the system is supporting ballistic spreading of energy,
the entanglement entropy of the system grows logarithmically (and not linearly) in time.
In Fig.~\ref{Fig:VNE_GS} we consider the reduced density matrix of the first $L/2$ sites of the system, $\hat \rho_{L/2}$,
and plot its von Neumann entropy $S (\hat \rho_{L/2}) = \text{Tr}[\hat \rho_{L/2} \log \hat \rho_{L/2}]$ as a function of time.
The growth is fully compatible with a logarithmic scaling $\sim \log (t J)$.
This behavior is only apparently contradictory and it is the result of the BA integrability of the model; as such, we expect it also in other integrable models.
We leave as an open question whether a possible CFT in curved space treatment would be possible for a setting like this~\cite{SDCV17, DuSC17}.
Note that in a generic situation with finite temperatures such an analysis would not be feasible because
entanglement entropy does not have a simple generalization to mixed states.

\section{Low-temperature transport and conformal behavior}\label{Sec:LowT}

We now proceed to a direct comparison of the low-energy transport properties in our parafermionic model with a  simple prediction obtained
by means of standard CFT techniques.
Specifically, at low temperatures, energy transport in systems which display a low-energy conformal invariance
can be captured by the following compact BD formula~\cite{BeDo12, BeDo15}:
\begin{equation}
 {\mathcal J}_E = \frac{\pi c}{12} \left( \beta_l^{-2} - \beta_r^{-2} \right), \label{Eq.BD}
\end{equation}
where $c$ is the central charge. We stress that, contrary to commonly studied frameworks, where $c=1$ or $c=1/2$, in the present case we have $c=4/5$.

\begin{figure}[t]
 \begin{center}
 \includegraphics[width=0.95\columnwidth]{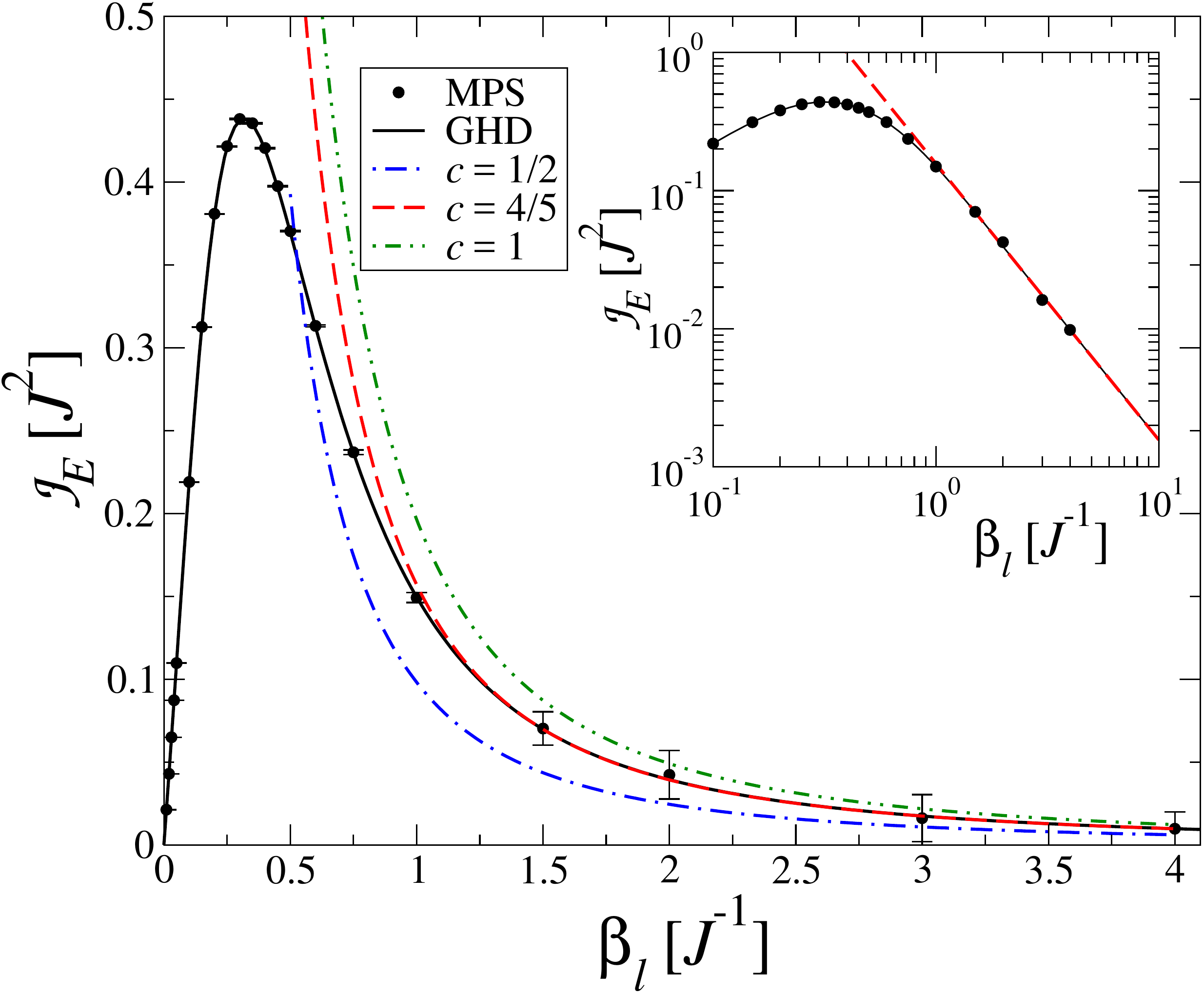}
 \end{center}
 \caption{Energy current ${\mathcal J}_E$ at the junction, as a function of $\beta_l$, for $\beta_r = 2 \beta_l$,
   according to the hydrodynamic theory (black solid line) and to numerical simulations (black circles).
   Simulations employ a maximal bond link $m = 300$ (see App.~\ref{App:LowT} for details).
   At low temperatures, we compare the data with the BD law for $c = 4/5$ (dashed red line).
   The more standard values $c=1$ (dotted-dashed green line) and $c=1/2$ (dotted-dashed blue line) are also shown.
   The inset clarifies the low-temperature scaling as $\beta_l^{-2}$ (straight red line).}
 \label{Fig:Current:Beta:c}
\end{figure}

The results of our analysis for the energy transport are summarized in Fig.~\ref{Fig:Current:Beta:c}, where we report steady-state values
of the energy current ${\mathcal J}_E$ flowing at the junction, according to GHD (black continuous lines), numerical simulations
with MPS (black circles), as well as the BD formula of Eq.~\eqref{Eq.BD} (dashed and dotted-dashed color lines).
For the sake of clarity and without lose in generality, we concentrate on the case $\beta_r = 2 \beta_l$. As such, temperature differences are significant and the data that we present go well beyond the possibilities of a linear-response theory.
Numerical data are obtained by performing the explicit evolution of the system in real time and extrapolating the long-time behavior of the energy current.
For $\beta_l>1.0 J^{-1}$, the energy current still displays non-negligible oscillations at the longest accessible times,
so that the extrapolation is susceptible to a non-negligible error (see App.~\ref{App:LowT} for details). 
The steady-state value is obtained by averaging the value of the current for the longest accessible times;
the error is estimated in a conservative way by taking the standard deviation of the points considered in the average
(see error bars in the figure).

We immediately recognize an excellent agreement between numerics and hydrodynamics. 
In order to quantitatively assess the validity of Eq.~\eqref{Eq.BD} in our case, we plot it not only for $c=4/5$,
but also for two other values of the central charge $c$ that are typically encountered in this kind of problems (namely, $1/2$ and $1$,
corresponding respectively to free-fermion and free-boson cases).
As expected, for large values of $\beta_l$, the system is well reproduced by the case $c=4/5$.
The quality of the agreement suggests that our analysis of numerical data overestimates the error performed in extrapolating the steady-state energy current.
Before concluding, we mention that Eq.~\eqref{Eq.BD} can be explicitly derived from the hydrodynamic theory using well-established techniques.
A more detailed discussion is reported in App. \ref{lowGHD}.

\section{Opposite temperatures}\label{Sec:OppT}

A peculiarity of Hamiltonian~\eqref{ham3} is that $\hat H$ and $-\hat H$ display a low-energy conformal limit with different central charges,
$c=4/5 $ and $c=1$, respectively. We now elaborate on the results presented in Sec.~\ref{Sec:DMRG:verification},
where we studied the PP for vanishing opposite temperatures, $\beta_r = - \beta_l \to \infty$, and argue that, in the limit
$|\beta_{l,r}| \gg J^{-1}$, we can study energy transport between two different CFTs.
We note that the models defined on spin-1/2 chains studied so far in the context of GHD do not offer this possibility, as in that case,
the low energy behavior of both $\hat H$ and $-\hat H$ are described by a $c=1$ CFT. 

The study of energy transport between models with different low-energy properties has been recently addressed in several contexts;
the results highlight a dependence on the specific nature of the junction through appropriate transmission coefficients, and as such are non universal.
Here we are effectively proposing a novel kind of junction that is integrable, since the two low-energy theories are connected
through an integrable model that interpolates between the two in energy space.
This approach has the clear disadvantage that, in the limit $|\beta_{l,r}| \to \infty$, the energy current is non-zero,
being equal to the energy current flowing between the ferromagnetic and antiferromagnetic ground states.
Yet, it allows for exact calculations of the long-time limits without invoking uncontrolled approximations or numerical estimates.

\begin{figure}[t]
\includegraphics[width=0.95\columnwidth]{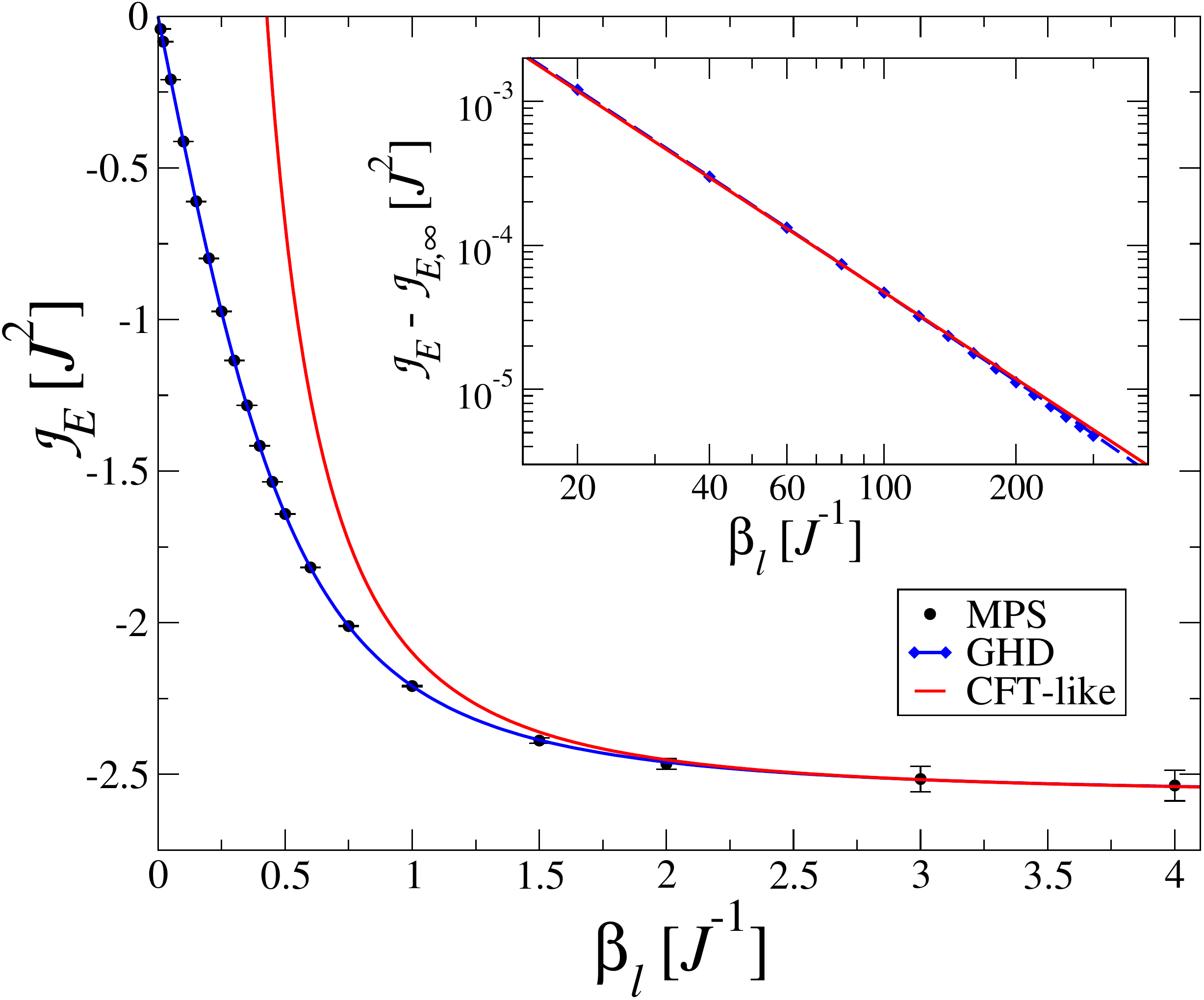}
\caption{Energy current ${\mathcal J}_E$ at the junction as a function of $\beta_l$, for $\beta_r = - \beta_l$, according to
  the hydrodynamic theory (blue solid line) and to numerical simulations with MPS (black circles).
  Simulations employ a maximal bond link $m=300$ (see App.~\ref{App:OppT} for further details).
  The red solid curve is the CFT-like formula in Eq.~\eqref{Eq.BDM}.
  In the inset we focus on large values of $\beta_l$, and show the hydrodynamic data (blue diamonds),
  the guessed formula in Eq.~\eqref{Eq.BDM} (red solid line) and the actual fit:
  ${\mathcal J}_E - {\mathcal J}_{E,\infty} = 0.485 \beta_l^{-2}-0.00016 \beta_l^{-1}$ (blue dashed line).
  The validity of the guessed CFT-like formula is only approximate.}
 \label{Fig:NegPos}
\end{figure}

Our results for the case $\beta_r = -\beta_l$ are presented in Fig.~\ref{Fig:NegPos}.
Steady values of the energy current ${\mathcal J}_E$ flowing at the junction are reported, both according to GHD and as computed with numerical MPS simulations. 
The steady values are obtained as in Sec.~\ref{Sec:LowT} (see App.~\ref{App:OppT} for details); as realized before, the agreement is excellent
also in this situation. 

It is now tempting to investigate whether a formula similar to~\eqref{Eq.BD} holds also in this case.
Based on formal analogies, we make the following intuitive guess:
\begin{equation}
  {\mathcal J}_E = {\mathcal J}_{E, \infty}+ \frac{\pi}{12} \left( c_l \beta_l^{-2} + c_r \beta_r^{-2} \right), \label{Eq.BDM}
\end{equation}
where $c_l$ and $c_r$ are the two central charges for the low-energy ($\beta_l \to + \infty$) and the high-energy ($\beta_r \to - \infty$)
conformal limit, respectively.
Here ${\mathcal J}_{E, \infty}$ denotes the non-universal energy current flowing in the limit $|\beta_{l,r}| \to \infty$,
which has been characterized in Sec.~\ref{Sec:DMRG:verification};
the supposedly universal behavior is sought in the fluctuations on top of it. 
Note that, since $\beta_l$ and $\beta_r$ have opposite signs, any non-zero value of $\beta_{l,r}^{-2}$ generates an energy current
flowing in the same direction.

We stress that, since we are connecting two halves with the maximal possible energy difference, the stationary state around the junction
will be far from any low-energy description and thus there is no good reason why the stationary energy current should obey
a simple relation like~\eqref{Eq.BDM}. Nevertheless, Eq.~\eqref{Eq.BDM} is exact for any non-interacting model
because there is no interaction between left- and right-moving excitations: their distribution in the stationary state only reflects the low-temperature behavior of the half they hail from.
As such, a deviation with respect to Eq.~\eqref{Eq.BDM} can be interpreted as a manifestation of interactions among the quasi-particle excitations.

At a first glance, the comparison of the hydrodynamic predictions with Eq.~\eqref{Eq.BDM} displays a surprising agreement,
as is apparent from Fig.~\ref{Fig:NegPos}.
A more careful inspection however shows that formula~\eqref{Eq.BDM} has only an approximate validity, as expected. 
Specifically, the behavior of ${\mathcal J}_E-{\mathcal J}_{E,\infty}$ predicted by the GHD for $\beta_l \geq 100 J^{-1}$
is well fitted by the function: $0.485 \beta_l^{-2}-0.00016 \beta_l^{-1}$. Note that the term scaling as $\beta_l^{-1}$,
which is not present in Eq.~\eqref{Eq.BDM}, is of order $10^{-4}$. Moreover, the expected prefactor of the $\beta_l^{-2}$ term,
that is $\frac{\pi}{12} (c_l + c_r) = \frac{\pi}{12} \frac 95 \approx 0.471 \ldots$, is compatible with the fitted behavior within few percents.
We thus conclude that the scaling proposed in Eq.~\eqref{Eq.BDM} is not exact, although rather accurate.
We leave as an open question the investigation of the reason for that.

\section{Conclusion}\label{Sec:Conc}

Motivated by the experimental interest that parafermions are raising in the condensed-matter and cold-atom communities, we investigated the energy transport in parafermionic chains. We employed the PP, according to which two semi-infinite chains are initialized at different temperatures and then let evolve in time. By choosing a specific parafermionic model that is integrable and solvable with BA, we developed a hydrodynamic description of the properties of the non-equilibrium steady state, and in particular of the energy current.
We validated the results with extensive numerical simulations based on time-dependent MPS.
The differences in the BA formulation of the parafermionic integrable chain with respect to more standard integrable models defined on spin-1/2 chains highlight the general validity of the GHD and its power as a tool for exploring the features of generic non-equilibrium steady states.

By studying the low-temperature energy transport of the model, we recovered the universal scaling described by the BD formula. For the first time, the formula was verified in a model with central charge different from $1$ and $1/2$, and namely $4/5$.
Motivated by the interest in the study of energy transport between models with different low-energy properties, we considered the case in which the two halves are initialized at opposite temperatures. As such, the half of the system at positive temperatures is close to the ground state, whereas the half of the system at negative temperatures is close to the maximally excited state (namely, the ground state of $-\hat H$).
We found that corrections to the limit $|\beta_{l,r}| \to \infty$ have a temperature dependence that is approximated by a universal function, whose origin cannot be explained by standard arguments based on conformal invariance.
Understanding the significance of this approximate validity is an open problem which will shed light on the role of interactions between the quasi-particles of the model.

\acknowledgments

We are grateful to B. Doyon for stimulating conversations.
This work was supported by  EPSRC Quantum Matter in and out of Equilibrium
Ref. EP/N01930X/1 (A.D.L.).
L.M. was  supported  by  LabEX  ENS-ICFP:  ANR-10-LABX-0010/ANR-10-IDEX-0001-02 PSL*.  
M.C. acknowledges support from the Quant-Era project "SuperTop" and the CNR-CONICET cooperation programme ``Energy conversion in quantum, nanoscale, hybrid devices''.
This work was granted access to the HPC resources of MesoPSL
financed by the Region Ile de France and the project Equip@Meso (reference ANR-10-EQPX-29-01)  of  the  programme  Investissements  d’Avenir  supervised  by  the  ANR. 
We acknowledge the CINECA award under the ISCRA initiative for the availability of high performance computing resources and support.

\appendix

\section{Details of the thermodynamic Bethe-Ansatz formulation \label{BADet}}

The Kernel matrix appearing in the BA equation~\eqref{fullBA} has the form $\Theta_{\mu, \nu}(\lambda, \mu) = \Theta_{\mu, \nu}(\lambda - \mu)$ with
\begin{align}
\label{ThetaExpl}
\Theta_{a,a}(\lambda) &= \Theta_{b,b}(\lambda) = \Theta_{c,c}(\lambda) = \Theta_{d,d}(\lambda) =   - \Theta_{a,d}(\lambda) = \nonumber \\ 
& = - \Theta_{b,c}(\lambda)= 2 \arctan \left[ \frac{\tanh \lambda }{\sqrt{3}} \right] , \\
\Theta_{a,b}(\lambda) &= \Theta_{c,d}(\lambda) = - \Theta_{a,c}(\lambda)= - \Theta_{b,d}(\lambda) =\nonumber \\ & = -2 \arctan \big( \sqrt{3} \tanh \lambda \big) , \\
\Theta_{a,m}(\lambda) &=  \Theta_{b,m}(\lambda) =- \Theta_{c,m}(\lambda) = -  \Theta_{d,m}(\lambda) = \nonumber \\ 
& =  2 \! \left[ \arctan \! \Big(\frac{\tanh \lambda }{2-\sqrt{3}} \Big) \! - \! \arctan \! \Big(\frac{\tanh \lambda }{\sqrt{3}+2} \Big) \!\right] \! . 
\end{align}
The matrix $\Theta_{\mu,\nu}$ is also almost symmetric, i.e. it gets a scale factor when transposed
\begin{equation}
\label{ThetaMatr}
\Theta_{\mu, \nu}(\lambda) = \left(\frac{1 + \delta_{\nu, m}}{1 + \delta_{\mu, m}}\right) \Theta_{\nu, \mu}(\lambda),
\end{equation}
which in matrix form can be rewtritten as $\Theta^t = S \Theta S^{-1}$,
with $S_{\mu,\nu}(\lambda, \mu) = \delta_{\mu,\nu} \, ( 1 + \delta_{\mu,e}) \, \delta(\lambda - \mu)$.

The source term $t_\mu(\lambda)$ has instead the form
\begin{align}
&t_{a}(\lambda) = 4 \arctan \left[ \big( \sqrt{3}+2 \big) \tanh (\lambda ) \right] ,\\
&t_{b}(\lambda) = -4 \arctan\left(\frac{\tanh \lambda }{\sqrt{3}+2} \right) \, ,\\
&t_{c}(\lambda) = 4 \arctan(\tanh \lambda ) - t_a(\lambda) , \\
&t_{d}(\lambda) = - 4 \arctan(\tanh \lambda ) - t_b(\lambda) ,  \\
&t_{e}(\lambda) = 4 \! \left[\arctan \Big( \frac{\tanh \lambda }{\sqrt{3}} \Big) \! - \!\arctan \big( \sqrt{3} \tanh \lambda \big) \right] \! .
\end{align}

\section{Additional information for the numerical simulations}

In this appendix we discuss some technical aspects of the simulations presented in the paper.

\subsection{Numerics presented in Sec.~\ref{Sec:DMRG:verification}}
\label{App:Accuracy}

\begin{figure}[t]
\begin{center}
 \includegraphics[width=0.85\columnwidth]{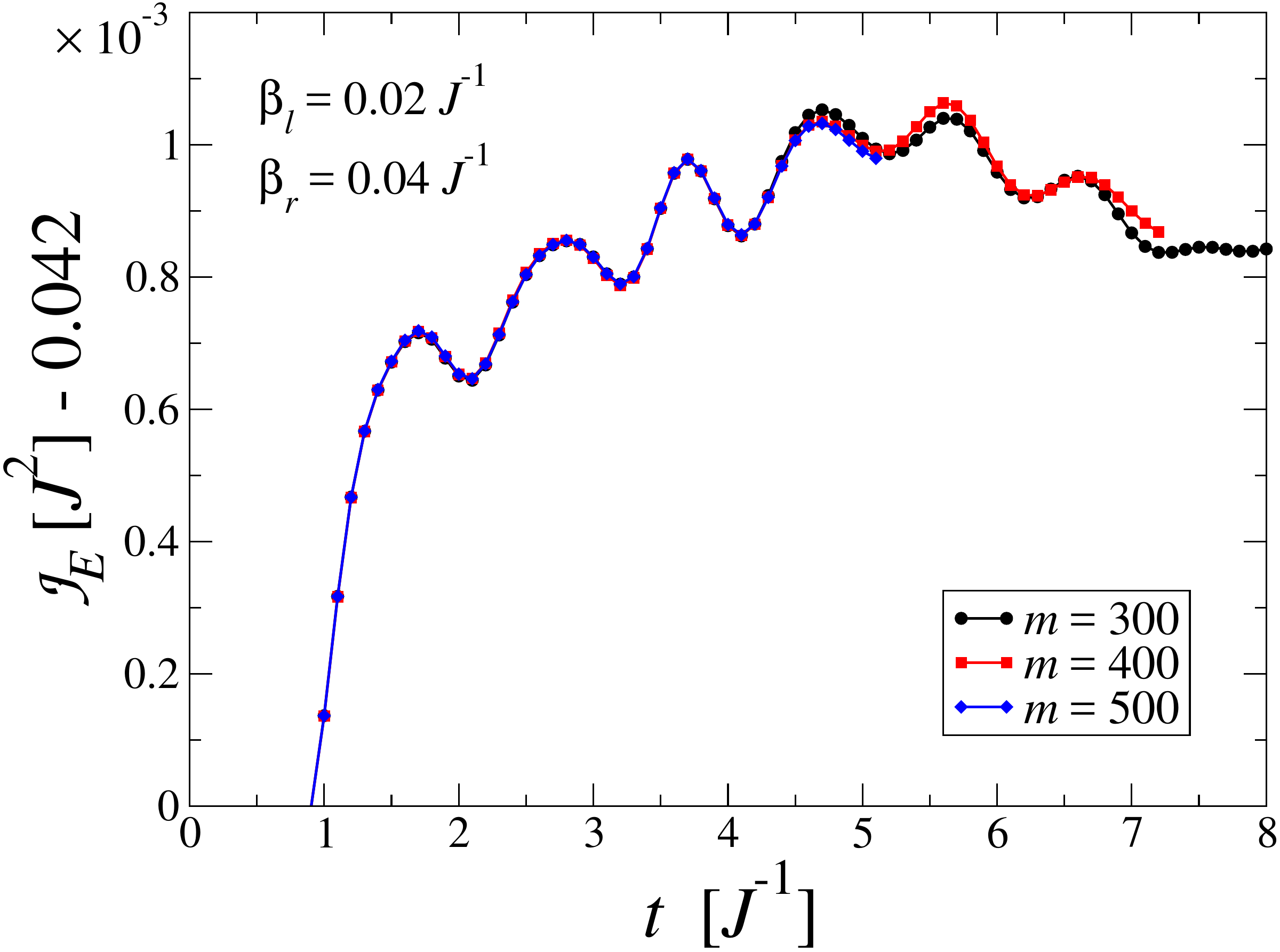}
\end{center}
\caption{Time evolution of the energy current for $\beta_l = 0.02$ and $\beta_r = 2 \beta_L$ for three different values of the maximal allowed bond link, $m$.
  Currents on the $y$-axis have been rescaled, so to highlight the tiny differences emerging when increasing $m$.}
 \label{Fig:App:Curr:Conv}
\end{figure}

In Fig.~\ref{Fig:App:Curr:Conv} we present the energy current ${\mathcal J}_E$ at site $x=0$ as a function of time, for the case
$\beta_l = 0.02 J^{-1}$ and $\beta_r = 2 \beta_l$. The three data sets ave been obtained allowing for three different maximal values
of bond link $m=300$, $400$ and $500$. The discrepancies can serve as estimates of the error committed in retaining the simulations
with the smallest bond link, $m=300$, and thus the lowest accuracy. Such error is estimated around $10^{-4} J^2$.
Note that an error of order $10^{-4} J^2$ does not affect the conclusions of the fitting procedure reported in Fig.~\ref{Fig:App:Hydro:Test:Thermal:2}.

\begin{figure}[t]
 \begin{center}
 \includegraphics[width=\columnwidth]{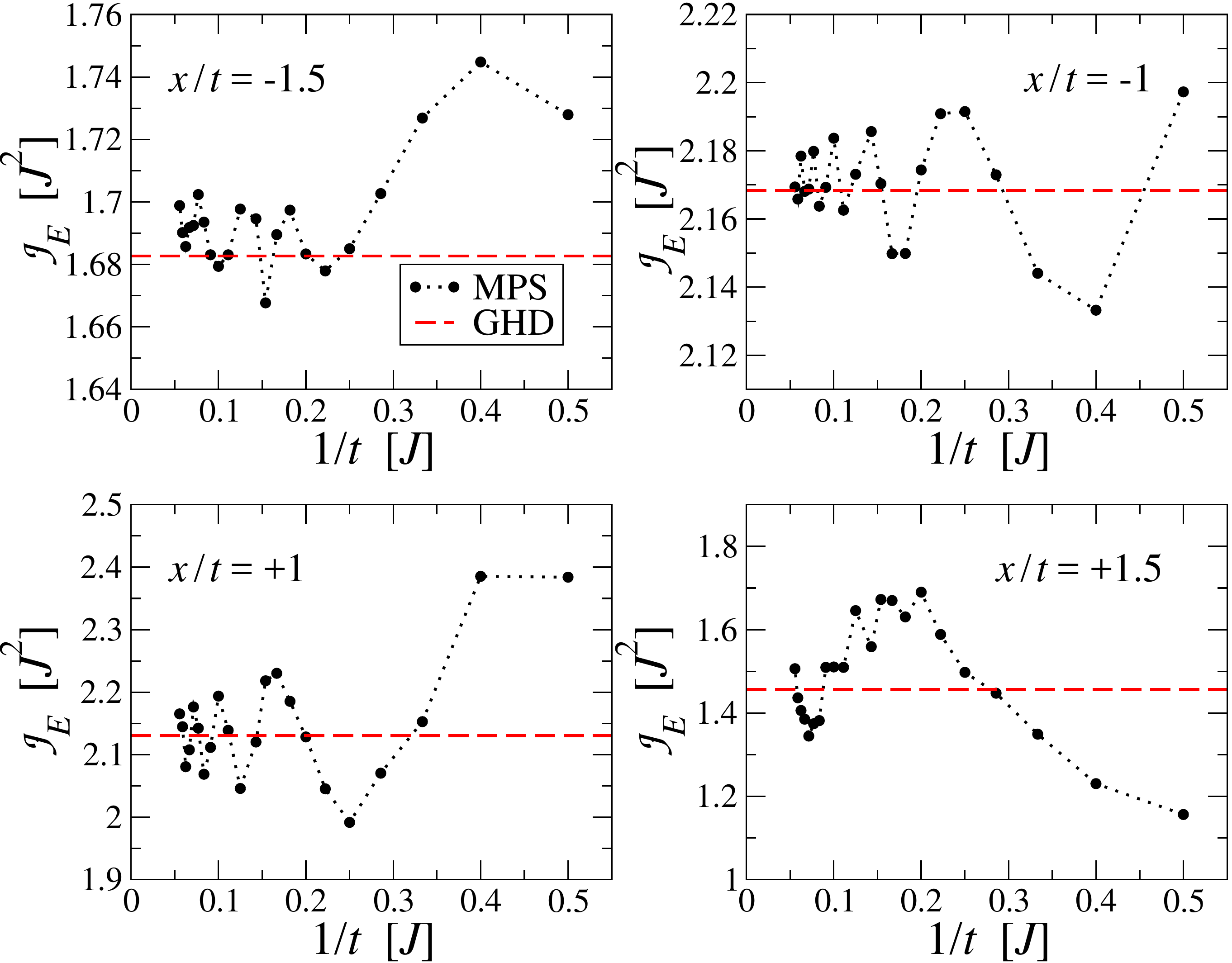}
 \end{center}
 \caption{
 Time-dependence of the data presented in Fig.~\ref{Fig:Hydro:Test:GS} for fixed values of $x/t = -1.5$, $-1.0$, $1.0$ and $1.5$. The horizontal red and dashed line represents the hydrodynamic prediction for the steady-state value.}
 \label{Fig:Hydro:Test:GS:2}
\end{figure}

In Fig.~\ref{Fig:Hydro:Test:GS:2} we study the time-dependence of the data presented in Fig.~\ref{Fig:Hydro:Test:GS}, considering in particular four representative values of $x/t$ and plotting the data as a function of time $t$. In all cases we observe an oscillatory behavior around the value obtained with the GHD.

\subsection{Numerics presented in Sec.~\ref{Sec:LowT}}
\label{App:LowT}

\begin{figure}[!h]
\begin{center}
\includegraphics[width=0.85\columnwidth]{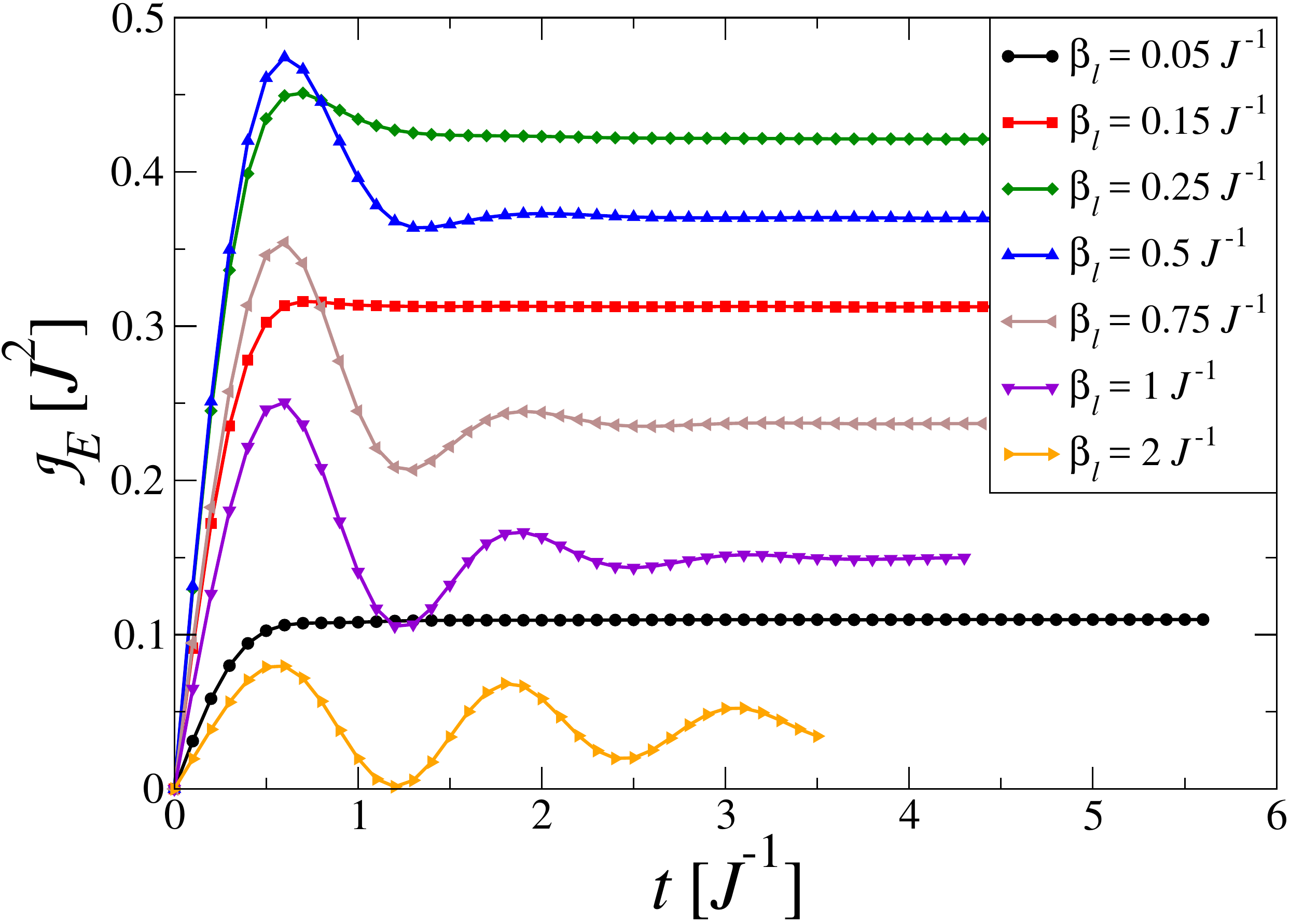}
 \end{center}
\caption{Time dependence of the energy current ${\mathcal J}_E$ at the junction for several values of $\beta_l$ and $\beta_r = 2 \beta_l$.
  Simulations employ a maximal bond link $m = 300$.}
 \label{Fig:App:Current:Time}
\end{figure}

The energy current ${\mathcal J}_E$ at the junction is plotted in Fig.~\ref{Fig:App:Current:Time} for several representative values
of $\beta_l$ (here, $\beta_r = 2 \beta_l$) as a function of time. 
For small values of $\beta_l$, at the longest accessible times any oscillatory behavior has been damped;
this is not the case for $\beta_l \geq 1.0 J^{-1}$. We thus average the data in the interval $[t_{\rm max} /2, t_{\rm max}]$
to extrapolate the steady value and take the standard deviation of the data to estimate the error committed in the procedure.

\subsection{Numerics presented in Sec.~\ref{Sec:OppT}}
\label{App:OppT}

\begin{figure}[!b]
\begin{center}
\includegraphics[width=0.85\columnwidth]{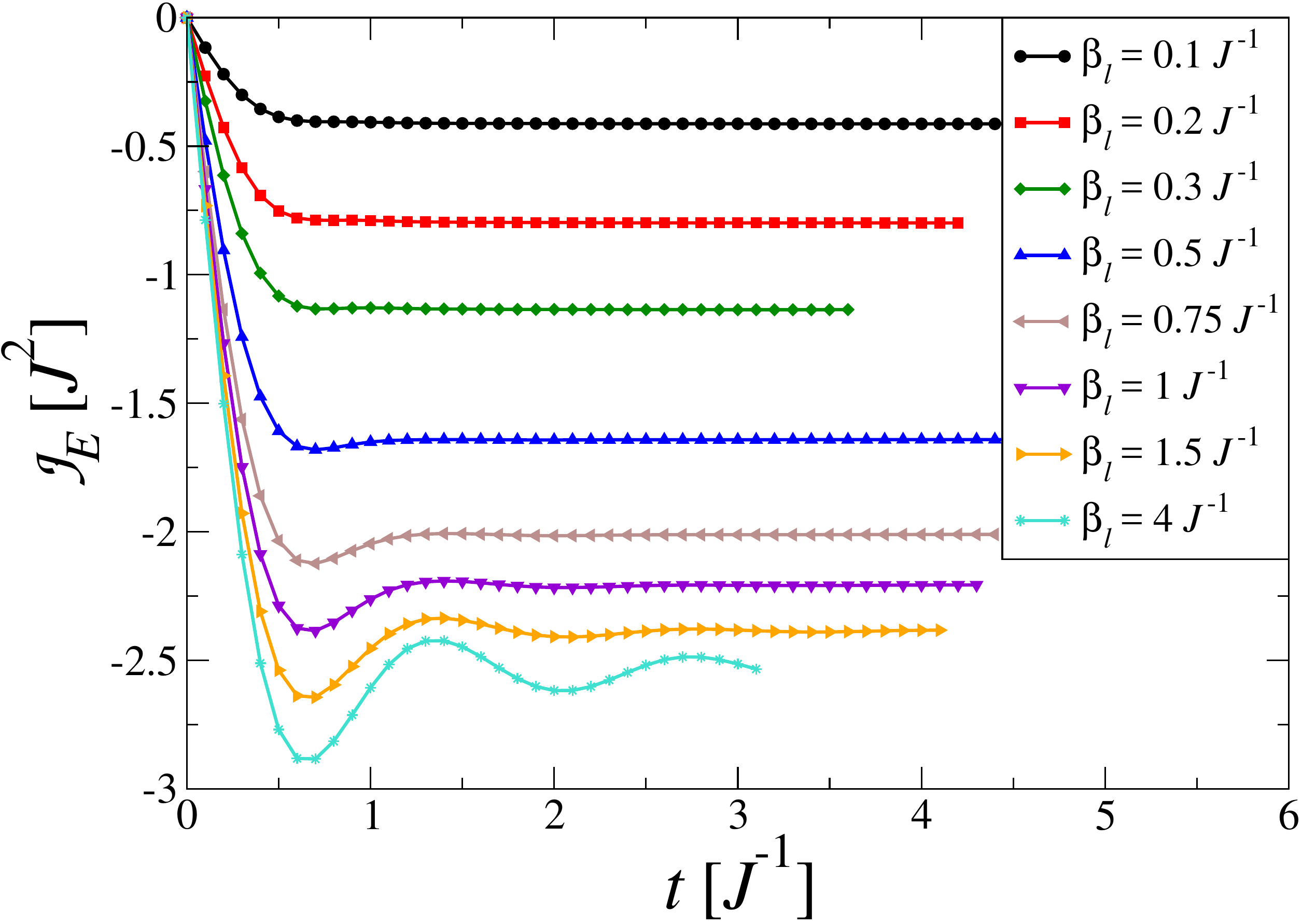}
 \end{center}
\caption{Time dependence of the energy current ${\mathcal J}_E$ at the junction for several values of $\beta_l$ and $\beta_r = - \beta_L$.
  Simulations employ a maximal bond link $m = 300$.}
 \label{Fig:App:Current:Time:NegPos}
\end{figure}

The energy current ${\mathcal J}_E$ at the junction is plotted in Fig.~\ref{Fig:App:Current:Time:NegPos} for several representative values
of $\beta_l$ (here, $\beta_r = - \beta_l$) as a function of time. 
For small values of $\beta_l$, at the longest accessible times any oscillatory behavior has been damped;
this is not the case for $\beta_l \geq 1.5 J^{-1}$. We thus average the data in the interval $[t_{\rm max} /2, t_{\rm max}]$
to extrapolate the steady value and take the standard deviation of the data to estimate the error committed in the procedure.

\section{Low temperature expansions from GHD \label{lowGHD}}

\begin{figure}[t]
\begin{center}
\includegraphics[width=0.85\columnwidth]{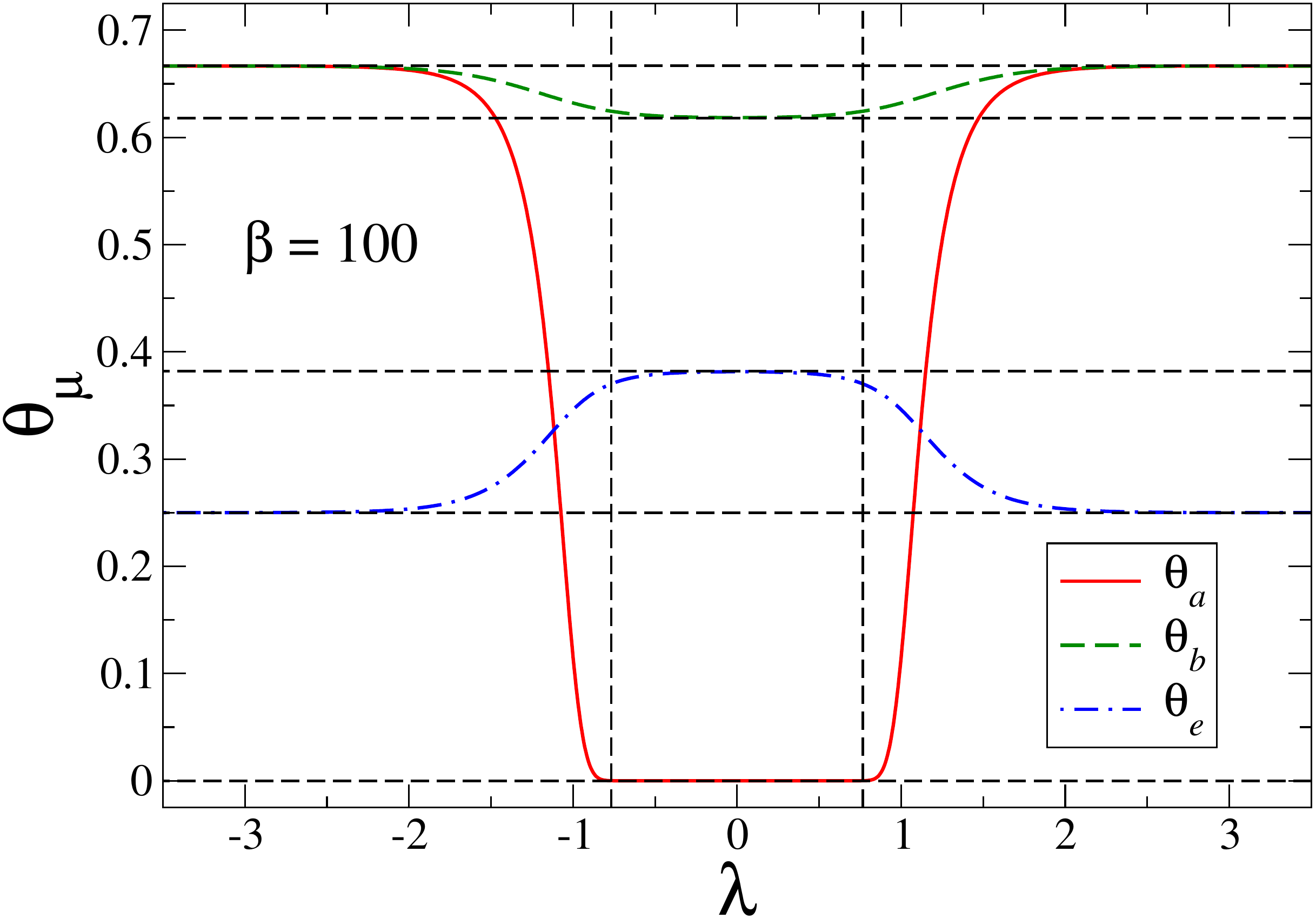} \vspace*{0.5cm} \\
\includegraphics[width=0.85\columnwidth]{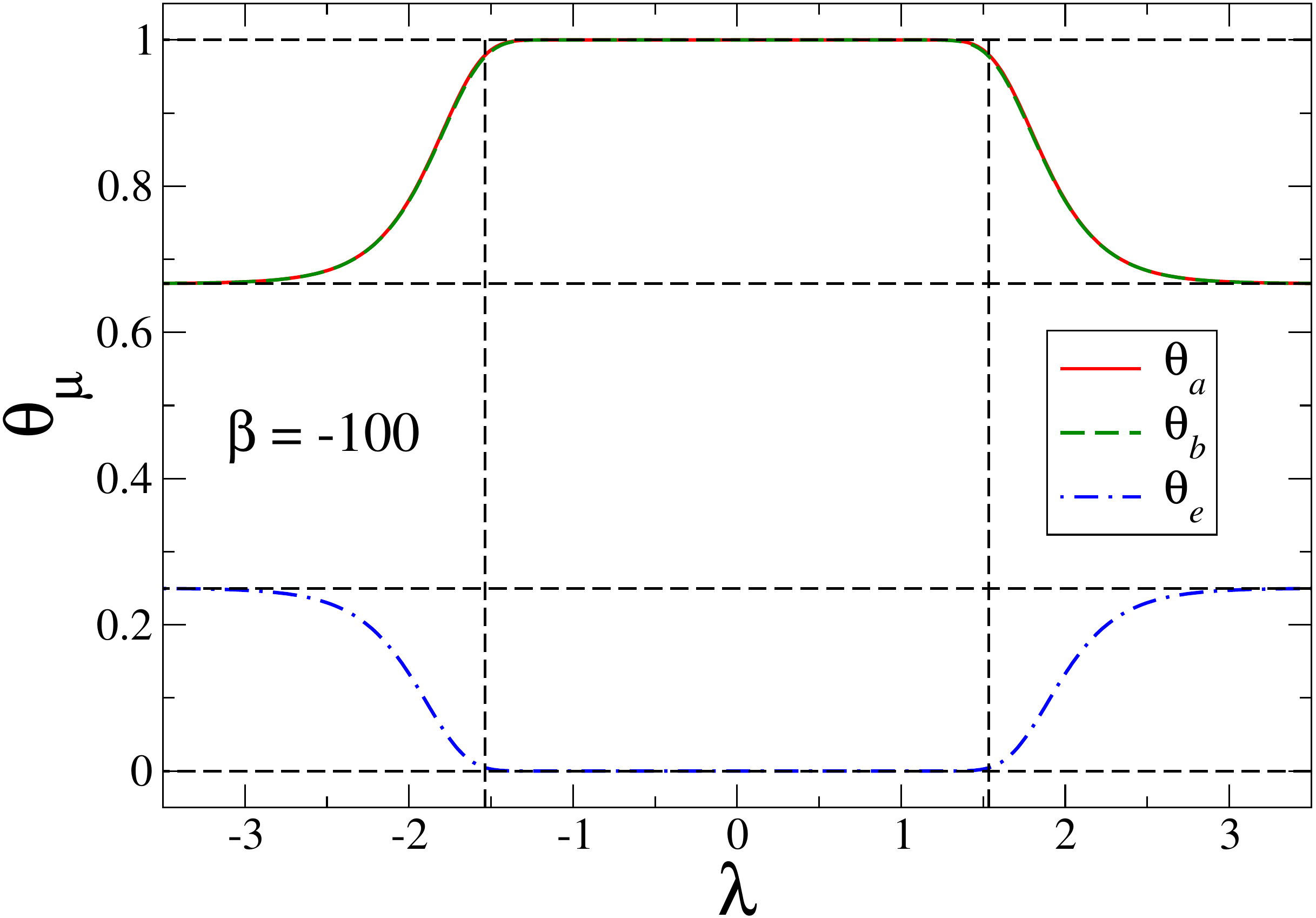}
 \end{center}
\caption{Behavior of the functions $\vartheta_\mu(\lambda)$ corresponding to a thermal state at a low positive temperature $\beta = 100$ (top),
  and a low negative temperature $\beta = -100$ (bottom).
  Horizontal dashed lines denote the asymptotic behaviors for $|\beta| \to \infty$ [Eqs.~\eqref{eq:asymp_beta}],
  and those for $\beta \to 0$ [Eqs.~\eqref{eq:asymp_beta0}].
  Vertical dashed lines correspond to $\lambda = \pm (\log \beta)/6$ (top)
  and $\lambda = \pm (\log |\beta|)/3$ (bottom).}
 \label{Fig:ExampleSmallT}
\end{figure}
\begin{figure}[t]
\begin{center}
\includegraphics[width=0.85\columnwidth]{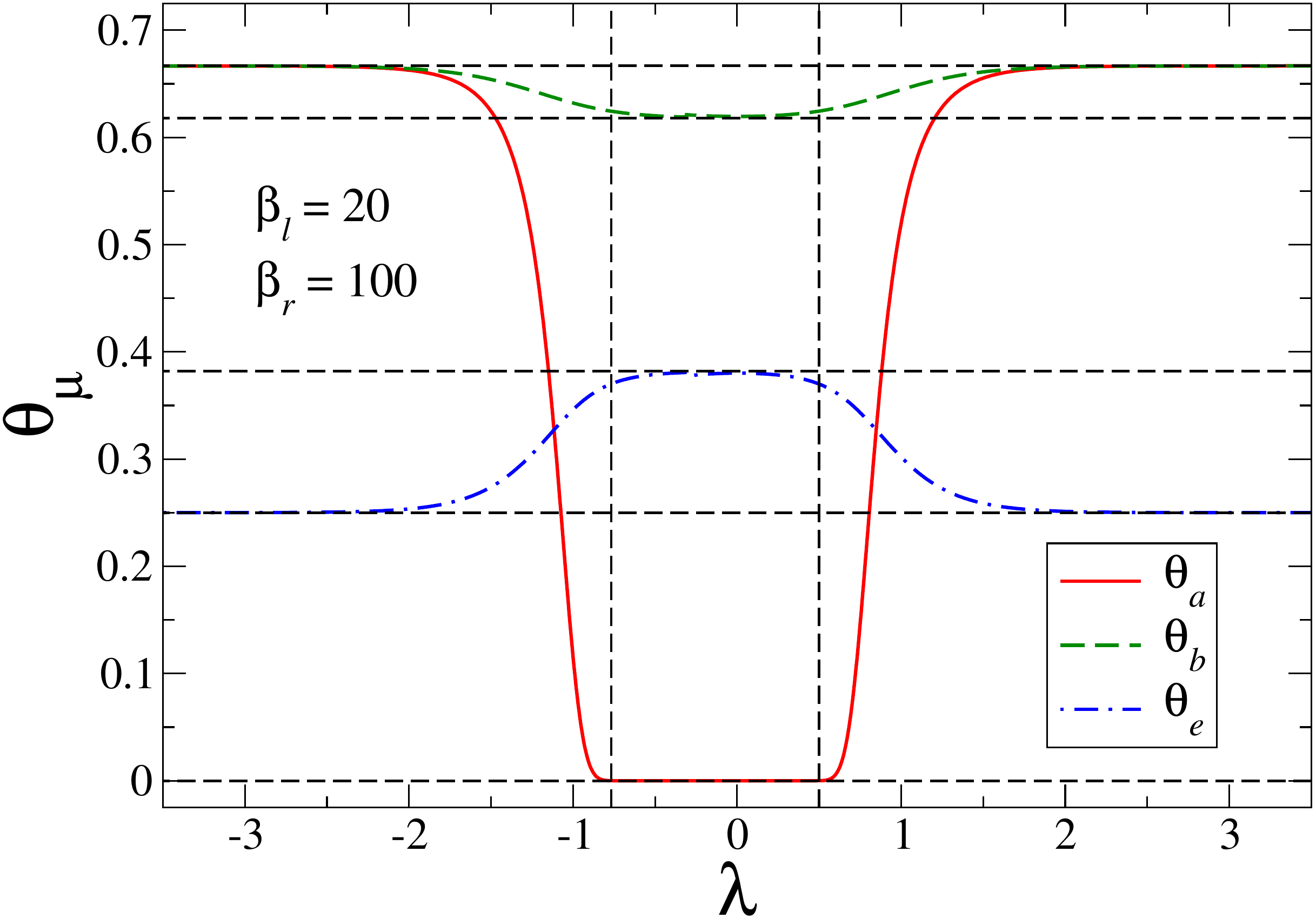} \vspace{0.5cm} \\
\includegraphics[width=0.85\columnwidth]{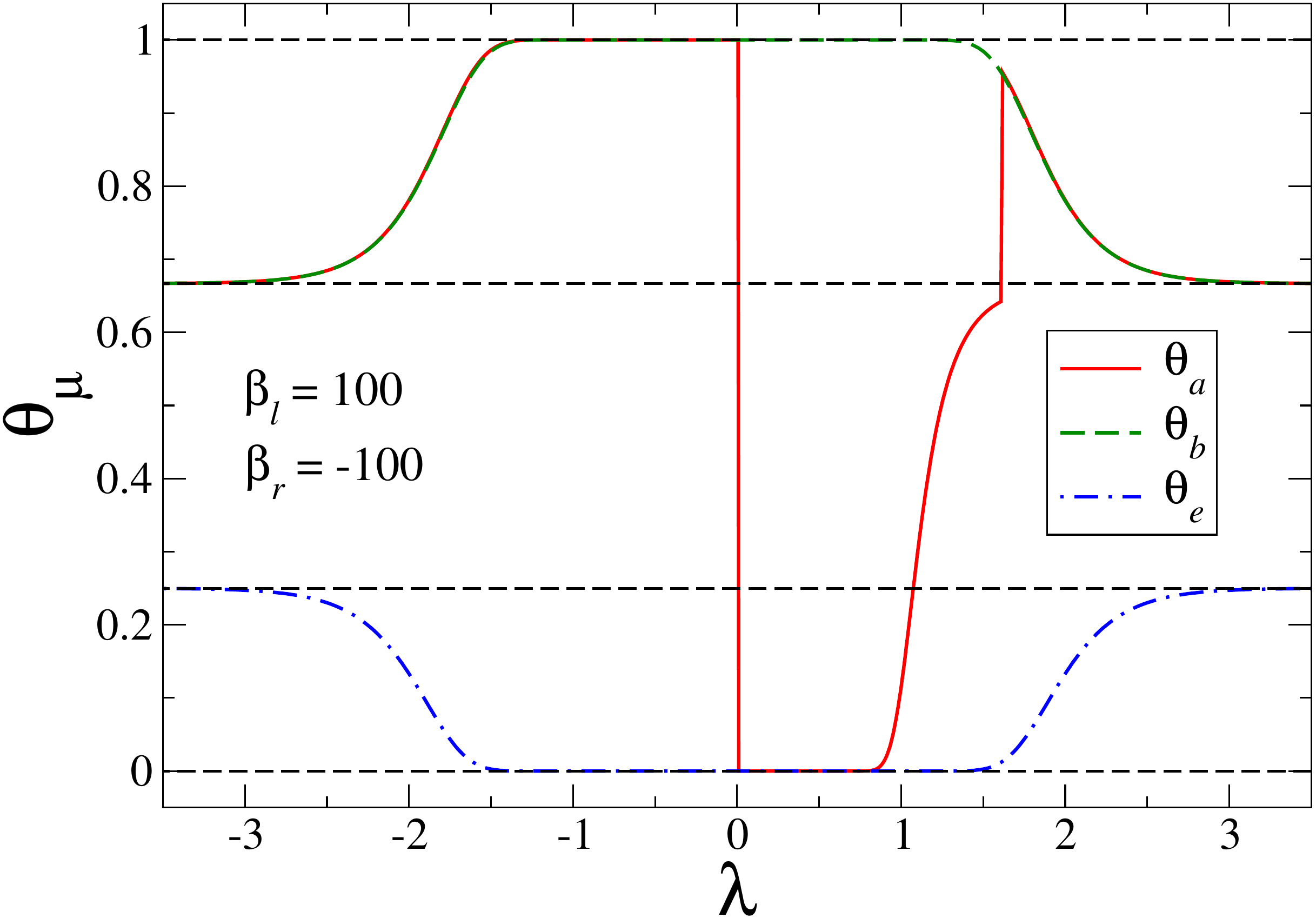}
 \end{center}
\caption{Behavior of the functions $\vartheta_\mu(\lambda)$ corresponding to the stationary state around the origin $x = 0$
  for the PP with $\beta_l = 20, \beta_r = 100$ (top) and $\beta_l = 100, \beta_r = -100$ (bottom).
   Horizontal dashed lines are the same as in Fig.~\ref{Fig:ExampleSmallT}.
  Vertical dashed lines in the top penal correspond to $\lambda = - (\log \beta_r)/6$ and $\lambda = (\log \beta_l)/6$.}
 \label{Fig:ExampleNess}
\end{figure}

In this section, we briefly discuss how Eq.~\eqref{Eq.BD} emerges from GHD and why Eq.~\eqref{Eq.BDM} is instead violated. We start noting that 
the solutions to Eq.~\eqref{TBAred} corresponding to the ferromagnetic ($\beta \to \infty$) and the antiferromagnetic ($\beta \to -\infty$)
ground state can be obtained explicitly. In particular, one finds that, in both cases, the solutions $\vartheta_{\mu}(\lambda; \beta \to \pm \infty)$
are independent of $\lambda$, and using \eqref{varthetofrometa} we get the values
\begin{equation}
  \label{eq:asymp_beta}
  \left\{ \begin{array}{l}
    \vartheta_{a} (\lambda, \beta \to \infty) = 0 ,\\ \vartheta_{b} (\lambda, \beta \to \infty) = \frac{\sqrt{5}-1}{2} , \\
    \vartheta_{e} (\lambda, \beta \to \infty) = \frac{3-\sqrt{5}}{2} ,
  \end{array} \right. \quad
  \left\{ \begin{array}{l}
\vartheta_{a} (\lambda, \beta \to- \infty) = 1 ,\\ \vartheta_{b} (\lambda, \beta \to- \infty) = 1 ,\\ \vartheta_{e} (\lambda, \beta \to -\infty) = 0 .
  \end{array} \right.
\end{equation}
Moreover, one can also obtain an explicit and constant solution for the infinite temperature case $\beta = 0$:
\begin{equation}
  \begin{cases}
    \label{eq:asymp_beta0}
\vartheta_{a} (\lambda, \beta = 0) = 2/3 \;,\\ \vartheta_{b} (\lambda, \beta = 0) = 2/3 \;,\\ \vartheta_{e} (\lambda, \beta = 0) = 1/4  \;.
\end{cases}
\end{equation}
For finite temperature, an analytic solution cannot be found. Nevertheless,
investigating the structure of \eqref{TBAred}, one deduces that 
 for large $|\beta|$, the functions $\vartheta_{\mu}(\lambda; \beta)$ have a rather simple structure~\cite{Kedem}, characterized by two flat asymptotic regimes. For $\beta \gg 1$, one has
\begin{align}
\vartheta_\mu(\lambda; \beta \gg 1) &= 
\begin{cases}
\vartheta_{\mu}(\lambda; \beta \to \infty) & \lambda \ll \frac{\log \beta}{6} \;,\\
\vartheta_{\mu}(\lambda; \beta = 0) & \lambda \gg \frac{\log \beta}{6}
\end{cases}\\
\vartheta_\mu(\lambda; \beta \ll -1) &= 
\begin{cases}
\vartheta_{\mu}(\lambda; \beta \to - \infty) & |\lambda| \ll \frac{\log |\beta|}{3} \;,\\
\vartheta_{\mu}(\lambda; \beta = 0) & |\lambda| \gg \frac{\log |\beta|}{3}
\end{cases}
\end{align}

In other words, the solution interpolates between the ground-state one at small $|\lambda|$ and the infinite temperature one at large $|\lambda|$, with a crossover scale which depends logarithmically on the inverse temperature $\beta$. An example is shown in Fig.~\ref{Fig:ExampleSmallT}. The correction of the thermal energy with respect to the ground-state value will only depend on the behavior of the functions $\vartheta_\mu(\lambda)$ around the crossover scale $\propto \log |\beta|$. In this way, with standard methods, one gets the universal corrections in agreement with CFT~\cite{takahashi, Kedem, Albertini1993}; consistently one obtains a vanishing thermal current as the contribution from $\lambda>0$ cancels exactly with the one at $\lambda < 0$.

With this information, we can briefly discuss the stationary state predicted by the GHD construction according to \eqref{selfpart} and we focus for simplicity on the case $x=0$. 
At low temperatures, the velocity $v_\mu^{[\vartheta]}(\lambda)$ has the same sign of $\lambda$ for $|\lambda| \gg 1$.
It follows that, for two low positive temperatures, the GHD solution is obtained by joining the thermal state at $\beta_l$ ($\lambda > 0$)
with the one at $\beta_r$ ($\lambda < 0$), see Fig.~\ref{Fig:ExampleNess} (top) for an example.
One can then compute the thermal current which results from the asymmetry from the positive and negative rapidities
and leads to~\eqref{Eq.BD} (see Ref.~\cite{bertinipiroli} for details). The case of two low and negative temperatures can be treated in a similar manner.

One could naively think that also when joining two opposite temperatures $\beta_l > 0, \beta_r < 0$ a similar construction could be applied. Indeed, this would be true if the dressing operation~\eqref{dressdef} had only a weak effect, as it trivially happens for free theories. On the contrary, when considering the junction between two thermal states close to the ferromagnetic and antiferromagnetic ground states, the dressing operation has a dramatic effect: for instance $v_b^{[\vartheta]}(\lambda) < 0$ and $v_e^{[\vartheta]}(\lambda) < 0$ for any value of $\lambda$. From~\eqref{selfpart}, this implies that the GHD solution for $\vartheta_{b}(\lambda)$ and $\vartheta_{e}(\lambda)$ equal the initial ones on the right, an example of this is given in Fig.~\ref{Fig:ExampleNess} (bottom). We stress that this is a rather counter-intuitive effect which is at the origin of the violation of~\eqref{Eq.BDM}. It is then surprising that the violation appears to be so small (see Fig.~\ref{Fig:NegPos}) and we leave this analysis to a future study.

\end{document}